\documentstyle[11pt,aas2pp4]{article}

\def\about            {\hbox{$\sim$}}

\def\comm#1           {{\tt (COMMENT: #1)}}
\def\dm               {\hbox{$\Delta m$}} 
\def\dt               {\hbox{$\Delta t$}} 

\begin{document}

\title{ A Description of Quasar Variability Measured Using Repeated SDSS 
and POSS Imaging }

Chelsea L. MacLeod\altaffilmark{\ref{Washington}},
\v{Z}eljko Ivezi\'{c}\altaffilmark{\ref{Washington}},
Branimir Sesar\altaffilmark{\ref{Caltech}},
Wim de Vries\altaffilmark{\ref{UCD}},
Christopher S. Kochanek\altaffilmark{\ref{OSU},\ref{CCAP}},
Brandon C. Kelly\altaffilmark{\ref{UCSB}},
Andrew C. Becker\altaffilmark{\ref{Washington}},
Robert H. Lupton\altaffilmark{\ref{Princeton}},
Patrick B. Hall\altaffilmark{\ref{York}},
Gordon T. Richards\altaffilmark{\ref{drexel}},
Scott F. Anderson\altaffilmark{\ref{Washington}},
Donald P. Schneider\altaffilmark{\ref{Penn},\ref{Penn2}}

\newcounter{address}
\setcounter{address}{1}
\altaffiltext{\theaddress}{University of Washington, Dept. of Astronomy,
Box 351580, Seattle, WA 98195, USA 
\label{Washington}}

\addtocounter{address}{1}
\altaffiltext{\theaddress}{Division of Physics, Mathematics and Astronomy, California Institute of Technology, Pasadena, CA 91125, USA\label{Caltech}}

\addtocounter{address}{1}
\altaffiltext{\theaddress}{University of California, One Shields Ave, Davis, CA 95616, USA 
\label{UCD}}

\addtocounter{address}{1}
\altaffiltext{\theaddress}{The Ohio State University, 140 West
  18th Avenue, Columbus, OH 43210, USA 
\label{OSU}}

\addtocounter{address}{1}
\altaffiltext{\theaddress}{The Center for Cosmology and Astroparticle Physics, The Ohio
  State University, 191 West Woodruff Avenue, Columbus, OH 43210, USA 
\label{CCAP}}

\addtocounter{address}{1}
\altaffiltext{\theaddress}{Department of Physics, University of California,
  Santa Barbara, CA 93106, USA
\label{UCSB}}

\addtocounter{address}{1}
\altaffiltext{\theaddress}{Department of Astrophysical Sciences, Princeton University, Princeton, New Jersey 08544, USA 
\label{Princeton}}

\addtocounter{address}{1}
\altaffiltext{\theaddress}{Department of Physics \& Astronomy, York
  University, Toronto, ON, M3J 1P3, Canada
\label{York}}

\addtocounter{address}{1}
\altaffiltext{\theaddress}{Department of Physics, Drexel University, 3141 Chestnut Street,
Philadelphia, PA 19104
\label{drexel}}

\addtocounter{address}{1}
\altaffiltext{\theaddress}{Department of Astronomy and Astrophysics, The Pennsylvania State University, 525 Davey Laboratory, University Park, PA 16802
\label{Penn}}

\addtocounter{address}{1}
\altaffiltext{\theaddress}{Institute for Gravitation and the Cosmos, The Pennsylvania State
University, University Park, PA 16802
\label{Penn2}}

\begin{abstract}
We provide a quantitative description and statistical interpretation
of the optical continuum variability of quasars. The Sloan Digital Sky  
Survey (SDSS) has obtained repeated imaging in five UV-to-IR
photometric bands for 33,881 spectroscopically confirmed quasars.
About 10,000 quasars have an average of 60 observations in each band
obtained over a decade along Stripe 82 (S82), whereas the remaining
\about25,000 have 2--3 observations due to scan overlaps. The observed
time lags span the range from a day to almost 10 years, and constrain
quasar variability at rest-frame time lags of up to 4 years, and at 
rest-frame wavelengths from 1000\AA\ to 6000\AA. We publicly release a
user-friendly catalog of quasars from the SDSS Data Release 7 that
have been observed at least twice in SDSS or once in both SDSS and the
Palomar Observatory Sky Survey, and we use it to analyze the ensemble
properties of quasar variability. Based on a damped random walk (DRW)
model defined by a characteristic time scale and an asymptotic
variability amplitude that scale with the luminosity, black hole mass,
and rest wavelength for individual quasars calibrated in S82, we can
fully explain the ensemble variability statistics of the non-S82
quasars such as the exponential distribution of large magnitude changes.  
All available data are consistent with the DRW model
as a viable description of the optical continuum variability of
quasars on time scales of \about 5--2000~days in the rest frame. We
use these models to predict the incidence of quasar contamination in
transient surveys such as those from PTF and LSST.  
\end{abstract}

\section{                    Introduction                     }

The optical continuum variability of quasars has been recognized since  
their first optical identification (Matthews \& Sandage 1963), and it has  
been proposed and utilized as an efficient method for their discovery 
(van den Bergh, Herbst, Prit-\\ chet 1973; Hawkins 1983; Hawkins \& Veron 1995; 
Ivezi\'{c} et al.\ 2004a; Rengstorf et al.\ 2006).  The observed
characteristics of the variability can then be used to
constrain the origin of their emission (e.g., Kawaguchi et al.\ 1998; 
Trevese, Kron \& Bunone 2001, and references therein). The variability
of quasars has typically been quantified using a structure function
(SF) analysis (e.g., Hughes et al.\ 1992; Collier \& Peterson 2001;
Bauer et al.\ 2009; Koz{\l}owski et al.\ 2010b; Welsh, Wheatley, \&
Neil 2011), where the SF is the root-mean-square (rms) magnitude
change (\dm) as a function of the time lag ($\Delta t$) between
measurements (similar to an auto-correlation function). It is fairly
well established that quasar variability properties depend on
physical properties such as the quasar luminosity, wavelength, time scale, and
the presence of radio emission. However, despite the considerable
observational effort invested over last few decades, many conflicting
claims about various correlations exist in the literature (see Giveon
et al.\ 1999 for a detailed discussion).  
 
The traditional method for studying variability has been to monitor a 
small, select sample of quasars over a long time baseline (e.g.,
Hawkins 2002; Giveon et al.\ 1999; Rengstorf et al.\ 2004). In this
case, it is possible to compute the SF for each quasar, which can
later be sample-averaged or studied individually. An alternative,
utilized in more recent studies based on Sloan Digital Sky Survey
(SDSS; York et al.\ 2000) and Palomar Observatory Sky Survey (POSS;
Minkowski \& Abell 1963) data, is to compute a single SF for all
quasars in a particular wavelength or luminosity range.  This
approach, mandated by the fact that typically only a few epochs were
available per object, only measures {\it ensemble} properties and 
{\it assumes} that all quasars selected from a narrow range of
physical properties vary in the same way. Nevertheless, this approach
enabled studies of quasar optical variability based on tens of
thousands of objects and several hundred thousand photometric 
observations, as well as explorations of the long-term variability
(Vanden Berk et al.\ 2004, hereafter VB04; Ivezi\'{c} et al.\ 2004c, hereafter I04;  
Wilhite et al.\ 2005; Mahabal et al.\ 2005; De~Vries, Becker, White,
\& Loomis 2005, hereafter dV05; Sesar et al.\ 2006, hereafter Ses06).  
For example, the size and quality of the sample analyzed by VB04
(two-epoch photometry for 25,000  spectroscopically confirmed quasars)
allowed them to constrain how quasar variability  in the rest frame
optical/UV regime depends upon rest frame time lag (up to  $\sim$2
years), luminosity, rest wavelength, redshift, the detection of radio
or X-ray emission, and the presence of broad absorption line systems.  
By comparing SDSS and POSS measurements for \about 20,000 quasars
spectroscopically confirmed by the SDSS, Ses06 constrained the optical
quasar variability on time scales from 10 to 50 years (in the
observer's frame). They report that there is a characteristic time scale  
of order 1 year in the quasar rest frame beyond which the SF flattens
to a constant value.  The SDSS has also facilitated both individual-
and ensemble-based approaches by providing a large multi-epoch sample
of quasars over the Northern Galactic Cap and well-sampled light
curves in the Southern Stripe 82 (S82) survey.  It is reassuring that
the two approaches lead to similar SFs, as discussed by dV05. A test
of this assumption is also described in MacLeod et al.\ (2008), who
show that indeed the {\it mean} behavior is the same. With such large
samples, the ensemble SF(\dt) slopes are well-constrained, and the
values suggest that accretion disk instabilities are the most likely
mechanism causing the observed optical variability (VB04; Kawaguchi et
al.\ 1998; see also Lyubarskii 1997). However, attempts to constrain
physical models using the ensemble SF are invalid as soon as one
realizes that the ensemble SF(\dt) is a weighted sum of individual
quasars with different structure functions (MacLeod et al.\ 2008).  

While studies have traditionally examined  ``non-parametric''
statistical measures of variability such as the SF, a
major challenge has been to describe the variability of individual
quasars in a compact way.  Recently, the introduction of a damped
random walk (DRW) model has provided a way to mathematically
characterize quasar light curves in terms of a characteristic
time scale ($\tau$) and an amplitude ($SF_{\infty}$) which are then 
correlated with the physical properties such as luminosity and black
hole mass. Kelly et al.\ (2009, hereafter KBS09) modeled a sample of
100 quasar light curves as a DRW and suggested that thermal
fluctuations driven by an underlying stochastic process such as a
turbulent magnetic field may be the dominant cause for the optical
flux fluctuations. Koz{\l}owski et al.\ (2010a; hereafter Koz{\l}10)
applied the DRW model to the well-sampled Optical Gravitational
Lensing Experiment (OGLE) light curves (Udalski et al. 1997; Udalski
et al. 2008) of mid-infrared-selected quasars behind the Magellanic
Clouds from Koz{\l}owski \& Kochanek (2009). Their analysis shows that
the DRW model is robust enough to efficiently select quasars from
other variable sources, despite the large surface density of
foreground Magellanic Cloud stars (see also Butler \& Bloom 2011;
MacLeod et al.\ 2011; Koz{\l}owski et al.\ 2011a). 

In MacLeod et al.\ (2010, hereafter Mac10), we applied the DRW model
to the light curves of \about 10,000 quasars in S82 and found a
correlation between $SF_{\infty}$ and black hole mass which is
independent of the anti-correlations with luminosity and wavelength
(see also Ai et al.\ 2010; Meusinger et al.\ 2011). We also found that $\tau$ increases with
increasing wavelength, remains nearly constant with redshift and
luminosity, and increases with increasing black hole mass (see also
KBS09; Koz{\l}10).  In Kelly et al.\ (2011), it was shown that a
similar stochastic model but with multiple time scales for a single
object can accurately reproduce the X-ray variability of active
galactic nuclei (AGN) and microquasars. An inhomogeneous accretion
disk model, where the temperature fluctuations throughout the disk are
driven by a DRW process, can explain the disk sizes derived from
microlensing light curves (see Morgan et al.\ 2010) while matching the
observed level of optical variability, and predicts SEDs which are in
better agreement with observations than standard thin disk models
(Dexter \& Agol 2011).  

One defining feature of the DRW model for a {\em single} quasar is
that it predicts a Gaussian distribution of magnitude differences \dm\
for a given \dt. On the other hand, the observed \dm\ distribution in
the optical for an {\em ensemble} of quasars observed at two times
separated by \dt\ deviates strongly from a Gaussian but is well fit by
an exponential distribution (I04).  This conflict represents an
important puzzle for understanding the DRW model and its applicability
to quasar light curves.  Also, the high likelihood of extreme values
of \dm\ has important implications for the interpretation of
observations of transients. For example, Vanden Berk et al.\ (2002)
reported the detection of an orphan gamma-ray burst afterglow based on
the 2.5 magnitude decrease in optical flux. Such a large flux change
was inconsistent with a quasar based on a Gaussian model for their
variability, but the source was nevertheless confirmed to be a highly
variable quasar (Gal-Yam et al.\ 2002). An accurate statistical
description of the two-epoch photometry for ensembles of quasars will 
be important for transient detection in large surveys, where quasars
represent a major contaminant.   

Our goal here is to produce a unified view of ensemble and individual 
optical variability in the context of the DRW model.  In this study,
we show that the differences in shape between the ensemble SF and the
DRW SF are well-explained by averaging over the properties of
individual quasars. We also show that the exponential distributions of  
magnitude changes for ensembles of quasars at fixed time lag are
naturally constructed by summing the intrinsically Gaussian
distributions of magnitude changes produced by individual
quasars. There are several residual issues which we discuss as part of
the comparison. An overview of the SDSS and POSS data used in this
study is presented in Section~\ref{data}. In Section~\ref{OV}, we
describe the observed ensemble quasar variability in terms of the SF
as a function of wavelength and time lag in the observer's frame.  We
then convert to rest-frame quantities and compare the data to a model
ensemble SF based on our previous DRW analysis of S82 light curves. In 
Section~\ref{LT}, we combine the constraints on short-term quasar
variability based on SDSS data with the constraints on long-term
variability derived from matching the SDSS and POSS catalogs. In
Section~\ref{FS}, we discuss the implications our results have on
transient identification, with a focus on future time-domain surveys
such as the Palomar Transient Factory and the Large Synoptic Survey
Telescope. Our results are discussed and summarized in
Section~\ref{Disc}.

\section{                   Data Overview                     }
\label{data}
In this section, we briefly summarize the relevant SDSS and POSS
data. We focus our description on the quasars with multiple
observations (similar to MacLeod et al.\ 2010; 2011).  

\subsection{  The Basic Characteristics of the SDSS Imaging Survey}
\label{phot}
The SDSS provides homogeneous and deep ($r < 22.5$) photometry in five
passbands ($u$, $g$, $r$, $i$, and $z$, Fukugita et al.\ 1996; Gunn et
al.\ 1998; Smith et al.\ 2002; Ivezi\'{c} et al.\ 2004b) accurate to
0.02 mag, of up to 12,000 deg$^2$ in the Seventh Data Release (DR7,
Abazajian et al.~2009). The DR7 sky coverage results in photometric
measurements for about 357 million unique objects. Astrometric
positions are accurate to better than 0.1 arcsec per coordinate (rms)
for sources brighter than 20.5~mag (Pier et al.\ 2003), and the
morphological information from the images allows robust star-galaxy
separation to \about 21.5~mag (Lupton et al.\ 2001). A compendium of
technical details about the SDSS can be found in Stoughton et al.\ (2002).

The SDSS offers an unprecedented photometric accuracy for such a large
scale optical survey. Not only are the photometric errors generally
small, but they are accurately determined by the photometric pipeline ({\em photo},
Lupton et al.\ 2001) and can be reliably used to estimate the statistical
significance of measured magnitude differences. This ability is of
paramount importance for a robust statistical study of variable objects. 
This error behavior is illustrated in Figure~2 in Ivezi\'{c} et al.\ (2003). 
Throughout our analysis, we assume the SDSS photometric errors to be
$\sigma_{phot}=0.018$~mag in $g$, $r$, and $i$ bands, and 0.04~mag in
$u$ and $z$ (see also Figure~1 in Sesar et al.\ 2007). The photometric
errors are assumed to be independent of magnitude due to the bright
sample limit of $i<19.1$ (see next section), and for simplicity when
accounting for photometric errors in the SF calculations. 

\subsection{           The SDSS Multi-epoch Data                  }

The SDSS imaging data are obtained by imaging the sky in six parallel 
scanlines, each 13.5 arcmin wide (a ``strip'' in SDSS
terminology). The six scanlines from two adjacent scans are then 
interleaved to make a filled ``stripe''. Because of the scan overlaps,
and because the scans converge near the survey poles, \about 40\% of
the sky in the northern survey is essentially surveyed twice.  In
addition, 290 deg$^2$ of the southern survey area lies along Stripe 82
(S82) and has been observed about 60 times on average to search for
variable objects and, by stacking the frames, to go deeper (Frieman et
al.\ 2008; Annis et al.\ 2011). This valuable subsample contains well-sampled light curves
for 9,258 spectroscopically confirmed quasars whose variability
properties are analyzed in Mac10, and it can be used to verify some of
the results inferred from the analysis of two repeated
observations. Overall, the SDSS has obtained multi-epoch data for 
\about 4,000 deg$^2$  of sky, with time scales ranging from 2 hours to
over 9 years, and with a wide range of Galactic latitudes extending 
all the way to the Galactic plane.  

\begin{figure}[t]
\epsscale{1}
\plotone{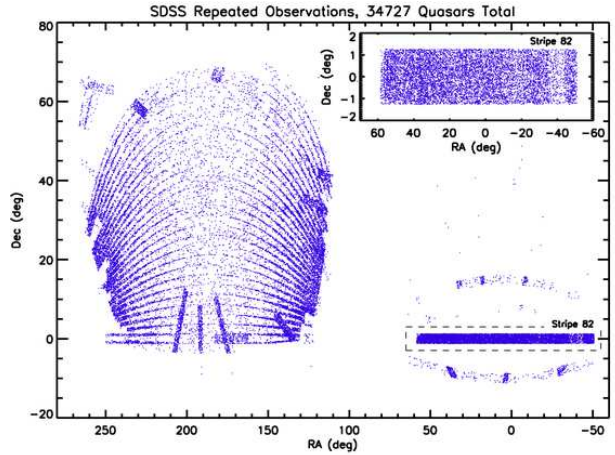}
\caption{\footnotesize Distribution of 33,881 DR7 quasars with at least 2 observations
on the sky in equatorial coordinates. The number of
observations increases towards the survey poles where the
``stripes'' overlap.  The inset shows a closer view
of the SDSS Stripe 82, which has \about 60 observations per object
(the density of quasars is non-uniform in S82 due to the increasing
Galactic contribution toward negative R.~A.). 
\label{RADec}}
\end{figure}

We define a quasar as any object listed in the SDSS catalog of
spectroscopically confirmed quasars (the ``DR7 Quasar Catalog'', Schneider 
et al.\ 2010). Its most recent fifth edition lists the SDSS DR7 BEST
photometry for 105,783 quasars from 9,380 deg$^2$. For a description
of the spectroscopic target selection for quasars, see Richards et
al.\ (2002). We note that the quality of photometry is good for the
objects in the DR7 Quasar Catalog. In total, there are 33,881
spectroscopically confirmed quasars with at least two observations. We
provide a catalog of all the SDSS repeated imaging of quasars 
online\footnote[12]{\scriptsize http://www.astro.washington.edu/users/ivezic/macleod/qso\_dr7/} 
(see the Appendix for a detailed catalog description).
The sky distribution of repeatedly imaged quasars is shown in Figure~\ref{RADec}. 

When we compare pairs of observations, we use the secondary imaging
observation with the largest time lag from the primary imaging
observation listed in the DR7 Quasar Catalog, for a total of two
observations per quasar. We define $\Delta m = m_2 - m_1$, where $m_2$
is the magnitude observed at a later epoch than $m_1$. We omit the S82
quasars since we do not wish to test our model on the same sample of
objects from which the DRW model parameters were derived.  The DR7
Quasar Catalog is spectroscopically  complete to $i<19.1$ in the
quasar region of color space. Therefore, unless otherwise noted, we
require that the primary observation for the remaining 24,627 quasars
has $i\leq 19.1$ in order to be consistent with the flux limit of the
quasar sample, for a total of 14,939 quasars. Given the SDSS limit of
$\sim$22~mag, we further restrict the magnitude differences in each
band to $|\dm|<3$~mag in order to reduce the amount of contamination
by poor photometry, for a total of \about 80,000 pairs of SDSS
measurements (summed over all bands).  
Finally, we adopt the redshifts and absolute magnitudes listed in the 
Schneider et al.\ (2010) spectroscopic quasar catalog, and the black
hole masses as measured from emission lines by Shen et al.\ (2011).
All magnitudes are corrected for Galactic extinction using
Schlegel et al.\ (1998). 

\subsection{SDSS--POSS Long-term Measurements }
While SDSS S82 obtained measurements with time differences of up to 10
years, longer time scales of up to 50 years can be probed by comparing
the SDSS and POSS catalogs. Ses06 have addressed the problem of large
systematic errors in POSS photometry by recalibrating several publicly
available POSS catalogs (USNO-A2.0, USNO-B1.0, DPOSS and GSC2.2). A
piecewise recalibration of the POSS data in 100 arcmin$^{2}$ patches
(one SDSS field) generally resulted in an improvement of photometric
accuracy (rms) by nearly a factor of two compared to the original data
(POSS-I magnitudes can be improved to $\sim$0.15 mag accuracy, and POSS-II
magnitudes to $\sim$0.10 mag accuracy). In addition to the smaller
core width of the error distribution, the tails of the distribution
become much steeper after the recalibration. These improvements are
mostly due to the very dense grid of calibration stars provided by
the SDSS, which rectifies the intrinsic inhomogeneities of Schmidt plates.

The much longer time lags between the Palomar Observatory Sky Surveys
(POSS-I and POSS-II) and the SDSS (up to $\sim$50 years) than spanned
by the available SDSS data make it easier to detect deviations of the
SF from a simple power-law.  dV05 and Ses06 compared the SDSS and POSS
data for over 10,000 quasars from the SDSS Data Release 2 in order to
constrain the long-term quasar variability.  As discussed by Ses06,
the dV05 and Ses06 measurements of the SF agree within a $1\sigma$ 
uncertainty of the Ses06 measurements. Here, we use results from both 
the dV05 and Ses06 studies. We also make use of data from the Digitized 
Palomar Observatory Sky Survey (DPOSS; Djorgovski et al.\ 1998) that 
overlap 8,000~deg$^2$ of sky from the SDSS Data Release 5. These DPOSS 
data were recalibrated following the procedure outlined by Ses06.
There are 81,189 SDSS DR7 
quasars with DPOSS observations in the $G$, $R$, and $I$ bands, and we
provide their 2-epoch photometry at the same website (see footnote 12). 
For our analysis in Section~\ref{LT}, we use the primary
SDSS observations as listed in the DR7 quasar catalog when comparing
SDSS--DPOSS magnitudes. We calculate synthetic
POSS $GRI$ magnitudes from the SDSS photometry, require $i<19.1$, and
impose various quality cuts following Ses06. In total, we have
56,732 SDSS--DPOSS \dm\ measurements.

\section{         Characteristics of Observed Variability         }
\label{OV}
For our SF analysis, we examine the behavior in both the
observer's and the quasar rest frames because the former is informative for
data interpretation (e.g., transients in large surveys) and the latter
constrains quasar physics. The SF behavior in the observer's frame is
discussed in Section~\ref{OF}, and we convert to rest-frame quantities in
Section~\ref{RF}.  We handle various trends with observed properties
by taking narrow ranges of the relevant quantities, in particular, the time
lag and wavelength in each frame. The luminosity, mass, and redshift
information are utilized in Section~\ref{RF}, where we test whether the
scalings with these physical parameters derived from individual light
curves in S82 can reproduce the ensemble variability of the
two-epoch sample. 

\subsection{Quasar Variability in the Observer's Frame}
\label{OF}

Assuming that the observed variability reflects the physics in the
accretion disk, the behavior in the observer's reference frame will be
a convolution of rest-frame variability over redshift, luminosity, and other
parameters. While this convolution will obscure important physical
scalings, quasar variability in the observer's reference frame is
still of major interest when interpreting survey data, in particular 
when distinguishing between quasars and other variable sources such as
transients, or in using variability to select quasars. Therefore, we
start by considering quasar variability in the observer's frame.  

\subsubsection{$\Delta m$ Distribution} 

\begin{figure}[t]
\plotone{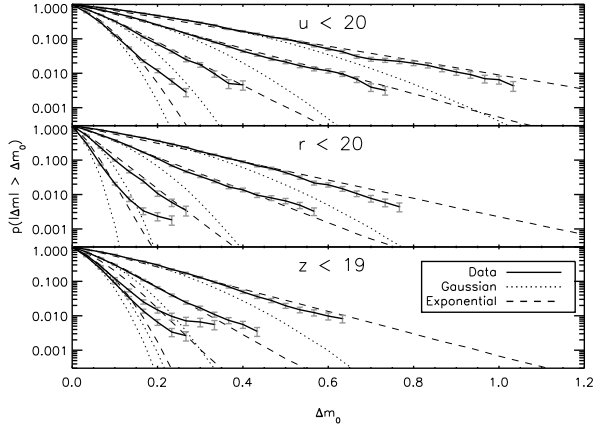}
\caption{\footnotesize The observed cumulative distributions of $urz$ magnitude
  differences as a function of time lag in the observer's frame.  We
  define $\Delta m = m_2 - m_1$, where $m_2$ is the magnitude observed
  at a later epoch than $m_1$.  The data (thick lines with Poissonian
  error bars) are only shown for sample sizes exceeding 10 data points
  (the probabilities become increasingly unreliable when below \about 0.003). 
  Going from the narrowest distribution to the widest, the time lags
  span 1--30~days, 50--150~days, 200--400~days, and 1400--3000~days.
  The thin lines show the predicted distributions based on Gaussian
  (dotted) and exponential (dashed) analytic functions with the same
  rms as the data for each time lag range, where the photometric errors are taken
  into account by adding a Gaussian component of width $\sigma_{phot}$ as
  defined in Section~\ref{phot}.
\label{cumulative}}
\end{figure}

Figure~\ref{cumulative} shows the cumulative distribution of \dm\ in 
the SDSS $u$, $r$ and $z$ bands for four different \dt\ ranges. 
The thin lines show the predicted distributions based on Gaussian
(dotted) and exponential (dashed) analytic functions with the same rms as
the data. The photometric errors are taken into account by adding a
Gaussian component of width $\sigma_{phot}=0.018$~mag in $r$ 
and 0.04~mag in $u$ and $z$. The exponential curves predict
a much higher probability of large magnitude changes than the Gaussian
curves, and the data follow this prediction (we only show the data for
bins with more than 10 points). The data points become increasingly
unreliable in the tails of each distribution due to small sample
sizes, and this fact may cause the large discrepancy with the
exponential curve for small $\dt$ in the $r$ and $z$ bands.

\begin{figure}[t!]
\plotone{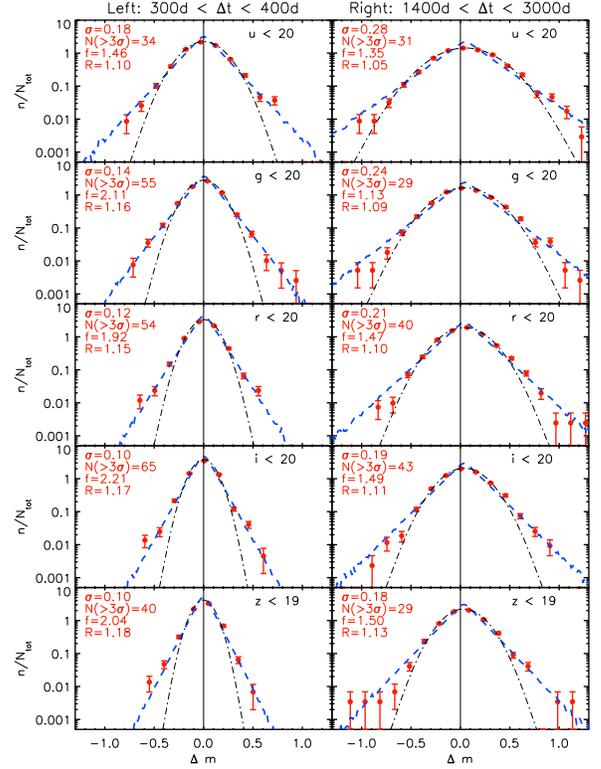}
\caption{\footnotesize The symbols show the distribution of measured magnitude 
differences, $\Delta m$, in the five SDSS bands, and for two narrow ranges  
of time lag, $\Delta t$, as indicated at the top. $n$ indicates the number 
of points in a bin divided by the bin width, and $N_{TOT}$ is the total 
number of points used for each histogram. The distribution width ($\sigma$)
calculated using Eq.~\ref{eq:iqr}, shown in each panel, is increasing with
time and decreasing with wavelength. The dot-dashed lines show
Gaussian distributions with a root-mean-square (rms) equal to $\sigma$,
and the dashed lines show exponential distributions, $\exp(-\Delta m/
\Delta_c)$, with $\Delta_c = {\rm rms}/\sqrt{2}$. Note that an exponential
distribution provides a better fit to the wings of the observed distributions. Each panel also
displays the number of measurements outside the $\pm 3\sigma$ range
($N>3\sigma$), their corresponding fraction ($f$) expressed in
percent (the expected value of this fraction for a Gaussian
distribution is 0.3\%, and for an exponential distribution 1.4\%), and
the ratio $R={\rm rms}/\sigma$.  
\label{dmHistObs}}
\end{figure}

Figure~\ref{dmHistObs} shows the differential \dm\ distributions for
each SDSS band and for two slices of observed-frame time lag. The
distribution width (i.e., the SF value) increases with time lag and
decreases with wavelength.  The wings of the distributions are also
closer to exponential (more accurately, a double-exponential or
Laplace dis- \\

\begin{figure}[t!]
\plotone{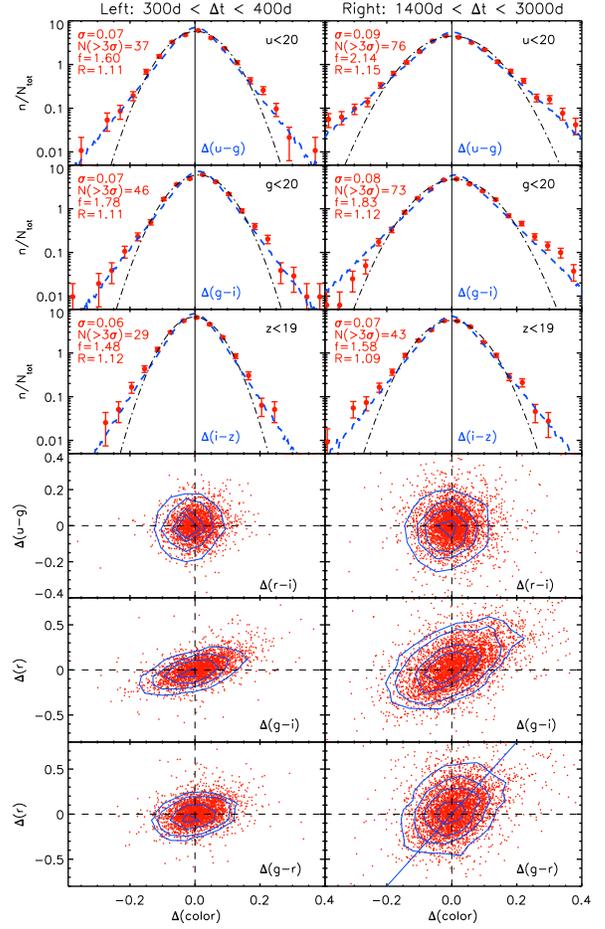} 
\caption{\footnotesize The top three rows are analogous to Fig.~\ref{dmHistObs},
  except that we show the changes in the $u-g$, $g-i$, and $i-z$ colors for
  two time lags in the observer's frame (left: 300--400 days, right:
  1400--1600 days). The fourth row shows the change of $u-g$ color as
  a function of the change of $r-i$ color, and the bottom two rows shows
  the change of the $r$ band magnitude as a function of the change of
  $g-i$ or $g-r$ color. The contours show regions containing 25\%, 50\%, 75\%
  and 90\% of the data points. The solid line in the bottom-right panel
  shows the slope found by Schmidt et al.\ (2011) using individual
  quasar light curves.
\label{dcHistObs}}
\end{figure}

\noindent tribution) than to Gaussian.  Figure~\ref{dcHistObs} shows
similar distributions for the changes in color for two slices in time
lag. The color changes are smaller than for the individual magnitudes
at the same time separation.  In the bottom six panels, the color
changes are plotted against each other as well as against the $r$-band 
differences. Although the scatter around these relationships is quite
large, quasars tend to get bluer as they brighten on average (bottom
four panels), in accordance with previous results (e.g., Giveon et al.\
1999). An analogous correlation was found recently by Schmidt et al.\
(2011), who analyzed individual light curves.  Their study showed the
slope $s_{gr}$ between the $r$-band and $g$-band variations, as in 
$r-\langle r\rangle=(s_{gr} + 1)(g-\langle g\rangle)+b$, is on average
around $-0.2$, indicating that quasars get bluer as they brighten. 
This value corresponds to a slope of 4 in the bottom-right panel of 
Figure~\ref{dcHistObs} (equating $m-\langle m\rangle$ to \dm), shown
by the solid line, which approximately follows the observed, two-epoch
distribution. 

\subsubsection{Structure Function}
\label{SF}
Next, we quantify the quasar variability using a SF analysis. Although
several definitions can be found in the literature, the SF essentially
measures the rms magnitude difference as a function of time lag
between magnitude measurements. We compute the structure function as
\begin{equation}
{\rm SF} = 0.74({\rm IQR})/\sqrt{N-1}, 
\label{eq:iqr}
\end{equation}
where IQR is the 25\% -- 75\% interquartile range of the $\dm$
distribution, and $N$ is the number of \dm\ values. This approach is
insensitive to outliers in the data which may result from poor data
quality and is especially effective at short time lags where the SF is
small.  This value is equivalent to the rms if the distribution is
Gaussian.  When stated, the photometric errors are taken into account
by subtracting $\sqrt{2}\sigma_{phot}$ in quadrature from the SF. Two
other definitions of the SF are found in the literature. Following
Bauer et al.\ (2009), we refer to them as SF$^{(A)}$ and SF$^{(B)}$:
\begin{equation}
{\rm SF}^{(A)}(\dt) = \sqrt{\langle \dm^{2}\rangle }
\label{sfa_eq}
\end{equation}
\begin{equation}
{\rm SF}^{(B)}(\dt) = \sqrt{\frac{\pi}{2}\left\langle \left|\dm\right|\right\rangle^{2} }.
\label{sfb_eq}
\end{equation}
For a Gaussian distribution, the ratio of the form adopted here to the 
other two forms is ${\rm SF}/{\rm SF}^{(A)} = {\rm SF}/{\rm SF}^{(B)} = 1$, while for an 
exponential (Laplace) distribution, ${\rm SF}/{\rm SF}^{(A)} = 1.03/\sqrt{2}$ and
${\rm SF}/{\rm SF}^{(B)} = 0.82$.  

\begin{figure*}[t!]
\epsscale{2}
\plotone{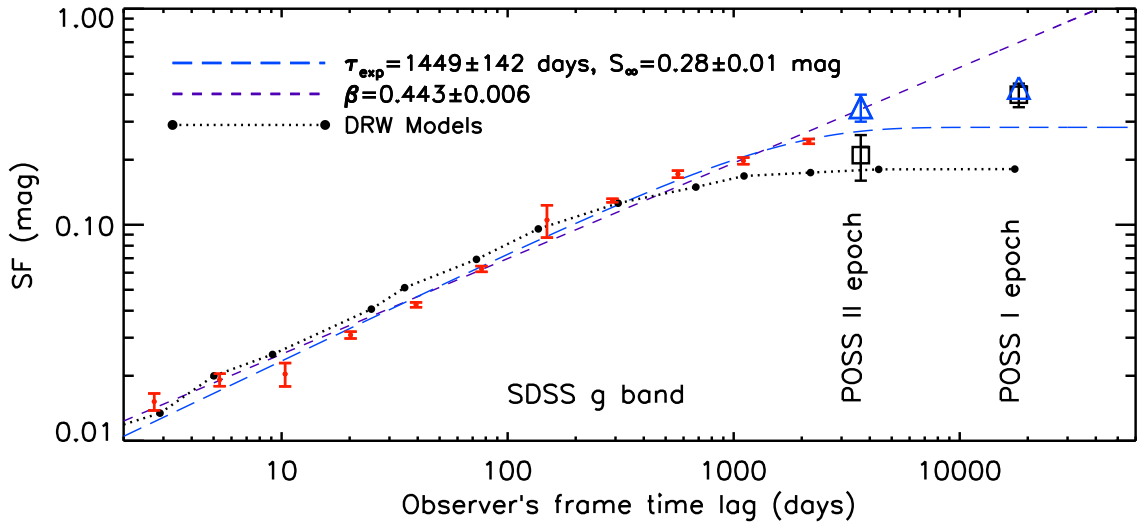}
\caption{\footnotesize The structure function for quasar variability measured in the 
  $g$ band and in the observer's frame (corrected for errors using 
  $SF = \sqrt{SF_{obs}^2 - 2\sigma_{phot}^2}$). The red data points
  show the SDSS measurements, and the data points at $\dt>3000$~days are inferred from
  comparing SDSS data and POSS I/II data, taken from Ses06 (open
  squares) and de Vries, Becker, and White (2003, blue triangles, which
  have similar errors as those from Ses06).  The
  short-dashed line shows the best power-law fit to the SDSS measurements
  alone, $SF \propto \Delta t^{\beta}$, and the long-dashed line shows
  a simultaneous fit to all data points, 
  $SF = S_{\infty}[1-\exp(-\dt/\tau_{\rm SF})]^{1/2}$. The best-fit
  parameters for these two fits are listed in the upper-left
  corner. The dotted line shows the prediction of the DRW model
  trained on S82, as described in Section~\ref{model}.    
\label{SFposs}}
\end{figure*}

Figure~\ref{SFposs} compares the SF for the combined SDSS and POSS
data over time scales ranging from 5~days to 50~years in the
observer's frame. The SF values shown up to $\dt\simeq 2000$~days are
computed using Eq.~\ref{eq:iqr}, with the photometric errors subtracted.  For variability
in the $g$ band with time lags of 10 and 50 years, we adopt 0.35 and
0.43~mag to represent the dV05 results, and 0.21 and 0.40~mag for the
Ses06 results, respectively. The errors in these estimates
are \about 0.05-0.10 mag.  The short-dashed line shows the best
power-law fit to the SDSS measurements alone (${\rm SF} \propto \dt^{0.44}$). It
overestimates the variability level  for time scales longer than 
several years. The long-dashed line shows the best-fit asymptotic 
function to all data points given by 
$S_{\infty}\, [1-\exp(-\dt/\tau_{\rm SF})]^{1/2}$, a common 
parametrization of the SF (e.g., Hook et al.\ 1994), with $S_{\infty}$
and $\tau_{\rm SF}$ as free parameters. The limiting value of the
best-fit overall structure function, $S_{\infty}=0.28\pm 0.01$~mag, is
probably systematically uncertain at a level of 0.02--0.03~mag due to
the discrepancies between the SDSS--POSS-I/II measurements (see
Ses06). The measured characteristic time scale, $\tau_{\rm SF} \sim
1400$~days, corresponds to \about 700~days in the rest frame and is
uncertain by $\sim$10--20\%.  While this analysis and the earlier ones
by dV05 and Ses06 provide strong evidence that the SF levels out on
long time scales, we cannot rule out a continuing but slower rise.

\subsection{             Quasar Variability in the Rest Frame         }
\label{RF}

\begin{figure}[t!]
\epsscale{1}
\plotone{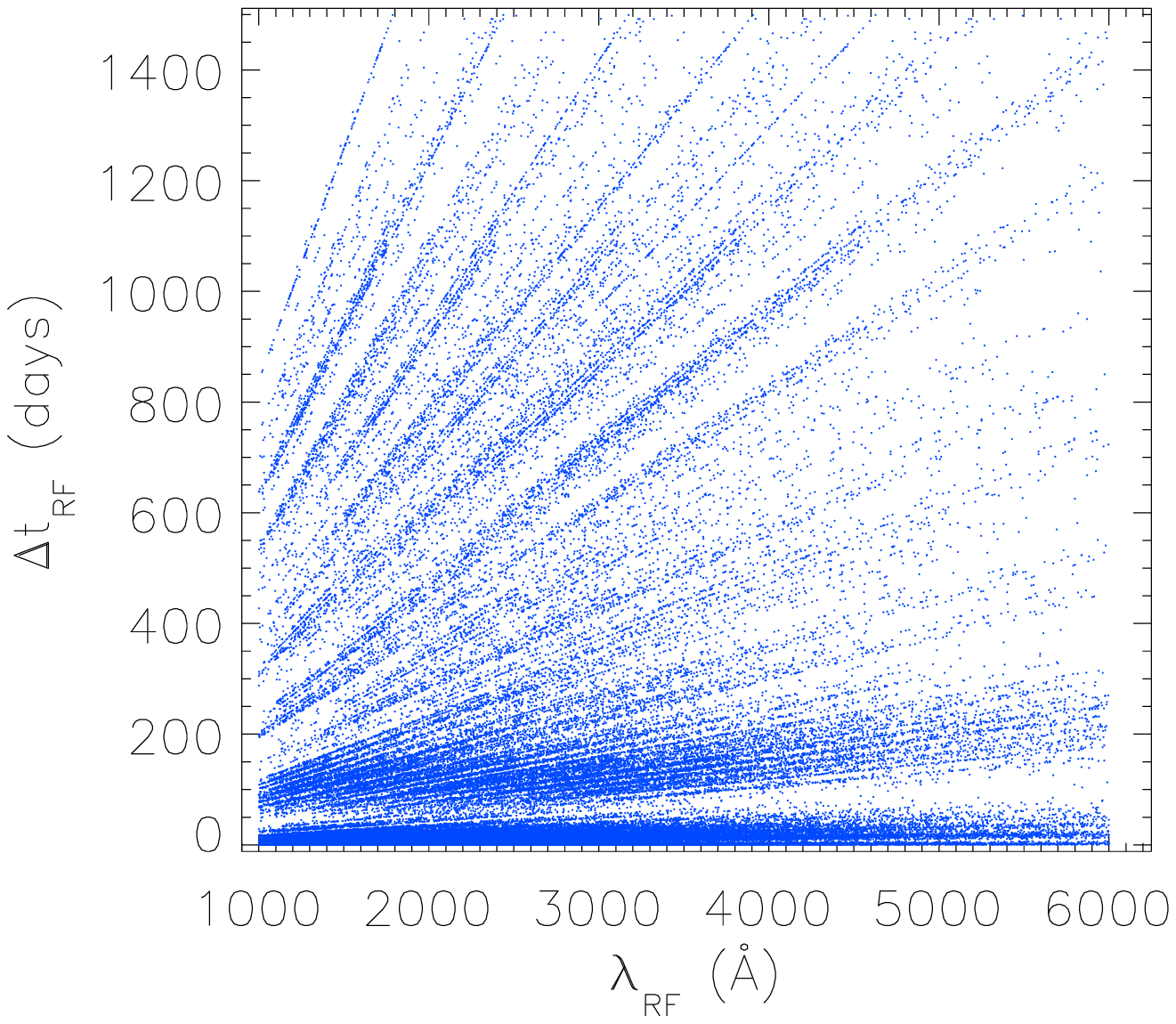}
\plotone{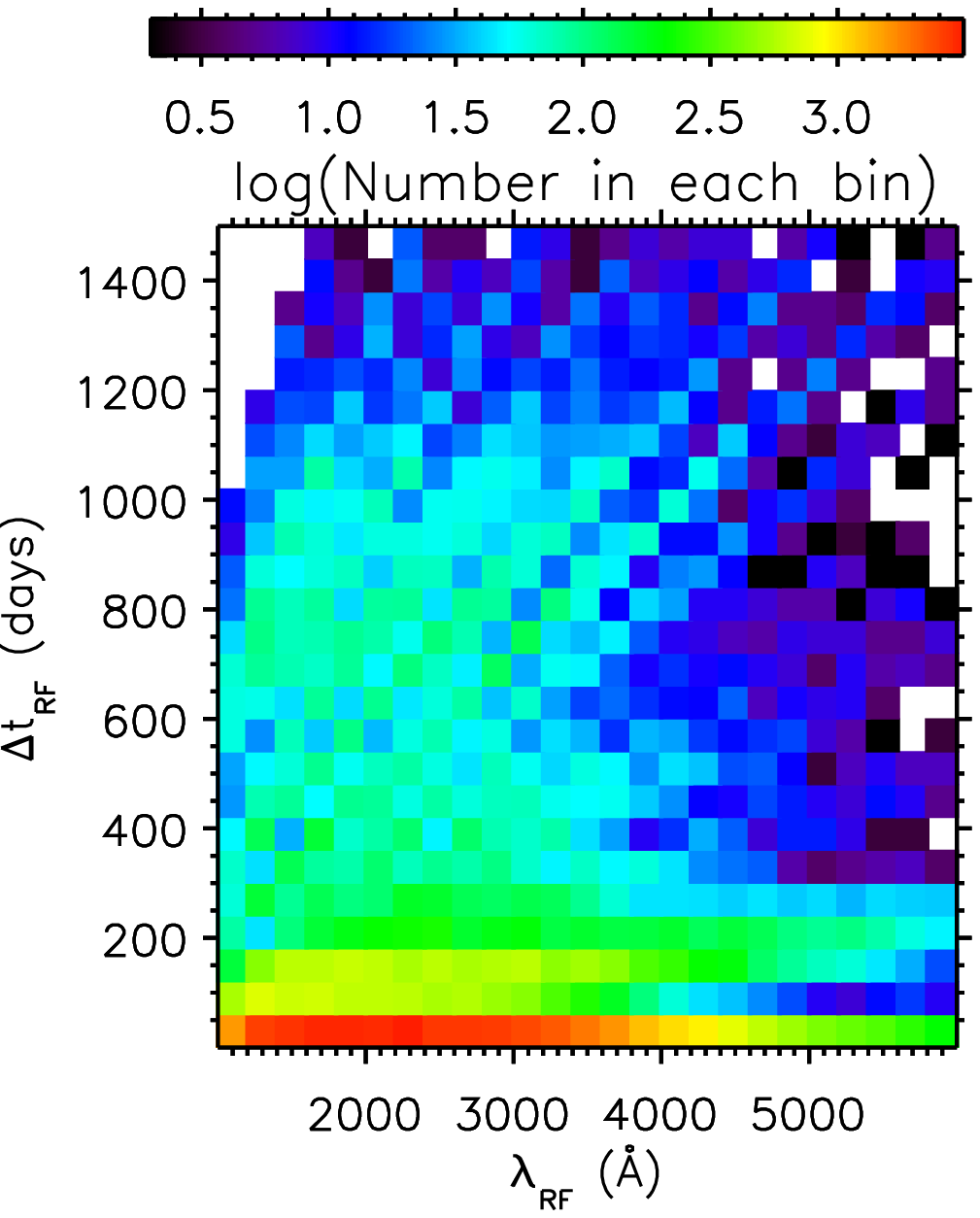}
\caption{\footnotesize (Top) The distribution of SDSS magnitude difference
  measurements in the plane spanned by rest-frame wavelength,
  $\lambda_{RF}$ and time lag, $\Delta t_{RF}$. The discrete
  distribution of $\lambda_o$ and $\Delta t_o$ in the observer's frame
  is spread along the lines $\Delta t_{RF} = k\, \lambda_{RF}$, with
  $k = \Delta t_o / \lambda_o$, and according to the sample redshift
  distribution [$\Delta t_{RF}= \Delta t_o / (1+z)$, $\lambda_{RF} =
  \lambda_o / (1+z)$]. The bottom panel shows the number of
  measurements in each bin on a log scale according to the legend at
  top. The white pixels contain fewer than 5 data points.
\label{dtRFlRF}}
\end{figure}

The redshift distribution of the quasars enables
one to map a discrete distribution of wavelengths and time differences
\dt\ in the observer's frame to a smoother distribution in the quasar rest frame.
Figure~\ref{dtRFlRF} shows the distribution of our 79,787 \dm\
measurements for $i<19.1$ objects in the rest-frame time difference
($\dt_{RF}$) and wavelength ($\lambda_{RF}$) plane.  Most (54\%)
observations have $\dt_{RF}<50$~days. The discrete distribution\footnote[13]{ 
  Note that a discretely-distributed
  observed \dt\ can lead to artificial correlations between the SF and
  $\dt_{RF}$.  For fixed $\dt_o$, longer $\dt_{RF}$ correspond to both
  lower redshift and, because of magnitude limits, lower
  luminosities. Since there is also an anti-correlation between
  variability amplitude and luminosity, this leads to a SF that
  increases toward higher $\dt_{RF}$. Therefore, caution must be taken 
  when binning in $\dt_{RF}$ using very sparse $\Delta t_o$ so that
  one does not mistake wiggles in the ensemble SF($\dt_{RF}$) for
  multiple intrinsic time scales, for example.} 
of $\lambda_o$ and $\Delta t_o$ in the observer's frame is spread 
along the lines $\Delta t_{RF} = k\, \lambda_{RF}$, with $k = \Delta 
t_o / \lambda_o$, and according to the sample redshift distribution 
[$\Delta t_{RF}= \Delta t_o / (1+z)$, $\lambda_{RF} = \lambda_o / (1+z)$].  
Here, we consider the shape of the \dm\ distribution as a function of
these quantities, the form of SF($\dt_{RF}$), and the dependence of the SF on
all physical parameters simultaneously, including the $i$-band absolute
magnitude $M_i$ (K-corrected to the rest frame) and the black hole mass
$M_{BH}$.

\subsubsection{Application of a DRW model}
\label{model}

We develop a model SF which reproduces the observed ensemble
variability using information derived from individual quasar light
curves in S82. Our model is motivated by the success of a DRW in
describing quasar light curves (KBS09; Koz{\l}10; Mac10), as well as
the detection of a turnover in the ensemble SF on long time scales,
suggesting a characteristic time scale for variability (e.g., Ses06;
Welsh, Wheatley, \& Neil 2011). We assume that the turnover is a
consequence of the characteristic time scale distribution measured
among individual quasars, and that the turnover in the ensemble SF 
corresponds to the average time scale, $\tau_{\rm SF}$. Suppose each 
quasar is described by its own DRW parameters, $\tau$ and  
${\rm  SF}_{\infty}$. Then, the ensemble SF is a weighted contribution 
of all the individual structure functions over their distribution in 
these parameters,   
\small
  \begin{equation}
      {\rm SF}(\Delta t) = \int d\tau d{\rm SF}_{\infty}{ d^2 n
	\over d\tau d{\rm SF}_{\infty} }
      {\rm SF}(\Delta t| \tau, {\rm SF}_{\infty})_{qso},
  \label{eq:ensSF}
  \end{equation}
\normalsize
where ${\rm SF}(\Delta t| \tau, SF_{\infty})_{qso}$ is the structure function at
time $\Delta t$ for a quasar with DRW variability  
\clearpage
\noindent parameters $\tau$ and ${\rm SF}_{\infty}$, given by 
\begin{equation}
  {\rm SF}(\Delta t| \tau, {\rm SF}_{\infty})_{qso} = {\rm SF}_{\infty}(1-e^{-|\Delta t|/\tau})^{1/2}.
\label{eq:sfdt}
\end{equation}
In particular, SF$(\Delta t)_{qso}$ is the expected standard deviation
of magnitude differences $\Delta m$ for a given quasar at a time lag 
$\Delta t$. SF$(\Delta t)_{qso}$ is related to the standard 
deviation in magnitudes ($\sigma_{m}$) at a given $\Delta t$ by
SF$(\Delta t)_{qso}=\sqrt{2} \sigma_{m}$, where the factor of
$\sqrt{2}$ results from subtracting two magnitudes.   

To build a model for the ensemble variability, we follow 
these steps for each quasar in the two-epoch sample:
\begin{enumerate}
\item Predict $\tau$ and SF$_{\infty}$ based on the quasar's physical parameters. 
\item Include the intrinsic scatter in $\tau$ and SF$_{\infty}$
  for quasars with similar physical parameters.
\item Estimate the SF value at the measured time lag $\Delta t_{RF}$ using Eq.~\ref{eq:sfdt}. 
\item Draw one model \dm\ value from a Gaussian distribution with a
  width set by SF$(\Delta t_{RF})_{qso}$, adding photometric noise if
  necessary.  
\end{enumerate}
We generally average over 1000 of these Monte Carlo models for the
distributions expected from the DRW model.  

In the first step, $\tau$ and SF$_{\infty}$ are estimated using the
scalings of the DRW parameters with the quasar's physical parameters
found in Mac10: 
\begin{eqnarray}
\log{f} = A + B\log\left(\frac{\lambda_{RF}}{4000{\rm \AA}}\right) + C(M_i+ 23) 
\nonumber \\ + D\log\left(\frac{M_{BH}}{10^9M_{\odot}}\right) + E\log(1+z),
\label{eq:form}
\end{eqnarray}
where $A = -0.51$, $B = -0.479$, $C = 0.131$, and $D = 0.18$ for
$f=$~SF$_{\infty}$ (in mag), and $A = 2.4$, $B = 0.17$, $C = 0.03$, 
and $D = 0.21$  for $f=\tau$ (in days). Note that the dependence on
redshift was found to be negligible in both cases ($E = 0$).  By
applying this model to the quasars in the large two-epoch sample, we are
also testing the accuracy of the scalings derived for the S82 quasars.  

\begin{figure}[t!]
\plotone{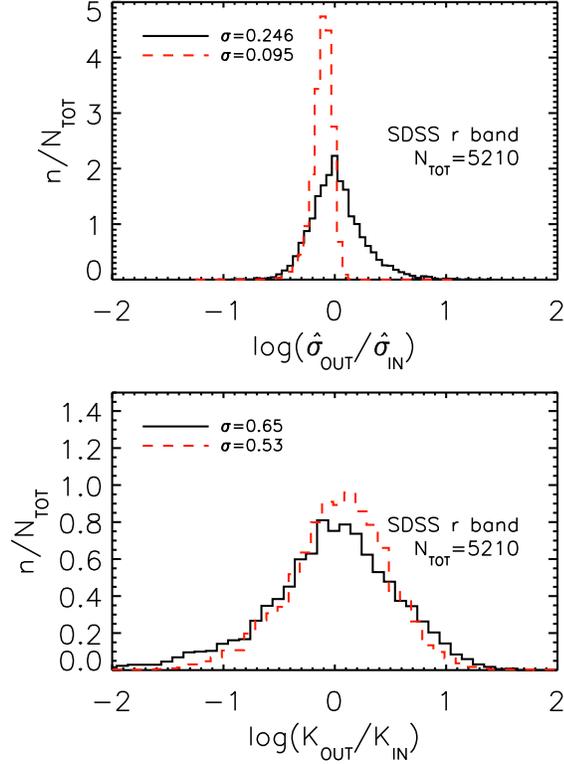}
\caption{\footnotesize Determination of fitting errors. The distributions of DRW
  parameters $\hat{\sigma}={\rm SF}_{\infty}/\sqrt{\tau}$ and  
  $K=\tau\sqrt{{\rm SF}_{\infty}}$ for Monte Carlo models of S82 quasar light
  curves are shown in the top and bottom panels, respectively.
  The solid, black curve in each panel shows the distribution of observed
  $\hat{\sigma}$ ($K$) values, which are used as inputs to generate 
  the light curves, normalized by the median value of
  0.16~mag~yr$^{-1/2}$ (200~mag$^{1/2}$~days). The red, dashed line
  shows the ratio between the best-fit parameters for the simulated
  light curves and their input values. The red histogram is narrower
  than the black histogram since the resulting best fit should be
  similar to the input value that generated the light curve. The
  contribution of fitting errors to the overall scatter of observed
  values is estimated to be the ratio of the variances ($\sigma^2$) of
  the red and black histograms. 
\label{fig:stotest}}
\end{figure}

In the second step, we account for the intrinsic scatter in $\tau$ and  
SF$_{\infty}$ for quasars with similar physical parameters. When
measured for individual quasar light curves, $\tau$ and SF$_{\infty}$
show scatter about their mean trends (Eq.~\ref{eq:form}).  The
magnitudes of these residuals are too large to be fully attributed to
measurement uncertainties (Mac10; Bauer et al.\ 2011). From
simulations (see Section 4.2 of Mac10), we estimated that the latter
statistical uncertainties account for only 70\%, 60\%, and 13\% of the 
scatter in $\tau$, SF$_{\infty}$, and 
$\hat{\sigma}={\rm SF}_{\infty}/\sqrt{\tau}$, respectively.   
The fitting errors are much smaller for $\hat{\sigma}$ because it is
well-constrained even for light curve lengths shorter than $\tau$. 
If we define $K=\tau\sqrt{{\rm SF}_{\infty}}$ as a variable orthogonal
to $\hat{\sigma}$ (in log space), its uncertainty due to fitting
errors contributes \about 82\% of its scatter. Figure~\ref{fig:stotest} 
compares the scatters in observed $\hat{\sigma}$ (top panel) and $K$ (bottom
panel) for S82 quasars to those for Monte Carlo models of the light curves. 

\begin{figure}[t!]
\plotone{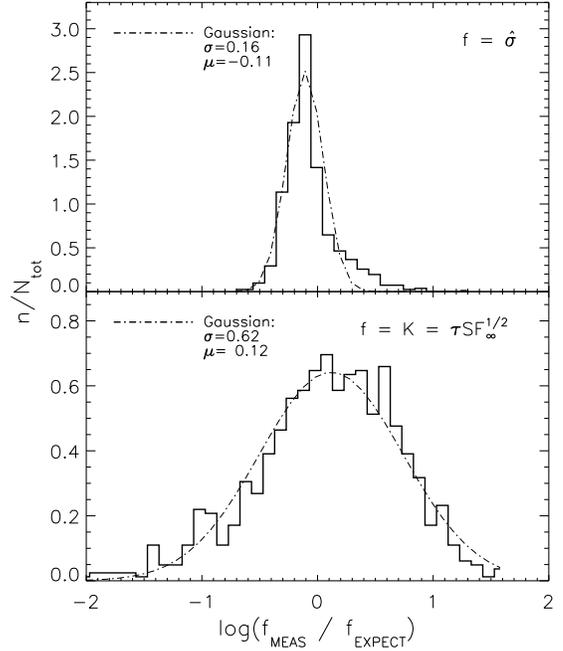}
\caption{\footnotesize Distributions of $\hat{\sigma}={\rm SF}_{\infty}/\sqrt{\tau}$ and 
  $K=\tau\sqrt{{\rm SF}_{\infty}}$ for S82 quasar light curves with
  $r<19$, normalized by the expected values from
  Eq.~\ref{eq:form}.   Each dot-dashed curve shows a Gaussian with
  unit area, with a mean ($\mu$) and rms ($\sigma$) as listed in the 
  legends. After removing the contribution from fitting errors ($F=13\%$
  and 82\% for $\hat{\sigma}$ and $K$, respectively), the intrinsic
  distribution widths are $\sigma_{int}=0.149$ and 0.26 dex, where
  $\sigma_{int}=(\sigma^2-F\sigma^2/100)^{1/2}$.  
\label{fig:spread}}
\end{figure}

Figure~\ref{fig:spread}
shows the distributions of $\hat{\sigma}$ and $K$ measured for the S82 
sample compared to their expected values from Eq.~\ref{eq:form}. We
find that the differences between the observed $\log \hat{\sigma}$
($\log K$) and the estimates from Eq.~\ref{eq:form} peak near zero
with an rms of 0.16 (0.62) dex. After taking the fitting errors into
account, the rms is reduced to 0.149 (0.26) dex, which we take as the  
intrinsic stochasticity. Therefore, in the second step, we take the
$\tau$ and ${\rm SF}_{\infty}$ estimated from Eq.~\ref{eq:form},
compute $\hat{\sigma}$ and $K$, add a random Gaussian deviate of width
0.149 (0.26) dex to each $\log \hat{\sigma}$ ($\log K$) value, and
then convert back to $\tau$ and ${\rm SF}_{\infty}$. This process
should provide a reasonable model for the intrinsic scatter. 

In the third step, the SF for a particular quasar at a given
$\Delta t_{RF}$ is estimated using Eq.~\ref{eq:sfdt}. In the fourth
step, a model $\Delta m$ value is drawn from a Gaussian distribution
with a standard deviation of SF$(\Delta t_{RF})_{qso}$, the Gaussian
distribution expected from the DRW model.  To account for
photometric errors, we further add a random Gaussian deviate of width
$\sqrt{2}\sigma_{phot}$ to each model \dm\ value (the $\sqrt{2}$
factor results from adding the photometric errors in quadrature). 
This procedure results in one model \dm\ value per quasar based on the
expectations from the DRW model. We then repeat this 1000 times for
each quasar in the sample. The resulting model for the ensemble SF is
then the rms width of this model \dm\ distribution otherwise
calculated in the same way as for the data (using the interquartile range).   

\subsubsection{Explaining the Exponential Tails of the \dm\ Distribution}

\begin{figure}[t!]
\epsscale{1}
\plotone{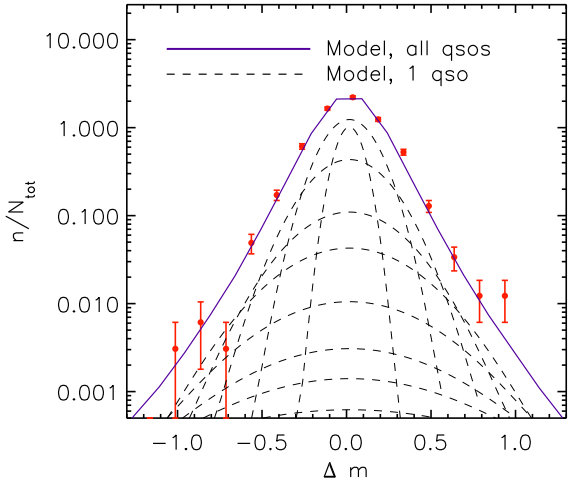}
\caption{\footnotesize The red data points with errors bars show the observed
  magnitude difference distribution for $365<\Delta t_{RF}<730$~days and 
  $2000{\rm \AA}<\lambda_{RF}<3000{\rm \AA}$. The solid line shows the
  expected distribution based on the DRW model. The dashed Gaussian
  curves show the expected distributions for 9 individual quasars with
  different values of SF$(\Delta t_{RF})_{qso}$ 
(and thus $\tau$ and SF$_{\infty}$; see Eq.~\ref{eq:sfdt}). Each
  curve has been convolved with a Gaussian noise component of width
  $\sigma_{phot}$ to account for photometric errors, and shifted by
  the median of the observed distribution.
\label{exptailmod}}
\end{figure}

In Figure~\ref{exptailmod}, the observed magnitude difference
distribution is shown for a narrow slice in  $\dt_{RF}$ and
$\lambda_{RF}$. The observed distribution is very similar to the model  
distribution (solid line), in which each quasar is assigned a model
$\Delta m$ value drawn from a particular SF$(\Delta t_{RF})_{qso}$, as 
described above. The exponential distribution for large $|\dm|$
results from a superposition of many Gaussians (dashed lines)
corresponding to different values of SF$(\Delta t_{RF})_{qso}$. The
range in SF$(\Delta t_{RF})_{qso}$ is caused by differing $\tau$ and 
SF$_{\infty}$ values, which can be attributed to a range in quasar 
luminosity and black hole mass plus intrinsic scatter.  
 
\begin{figure*}[p!]
\epsscale{2}
\plotone{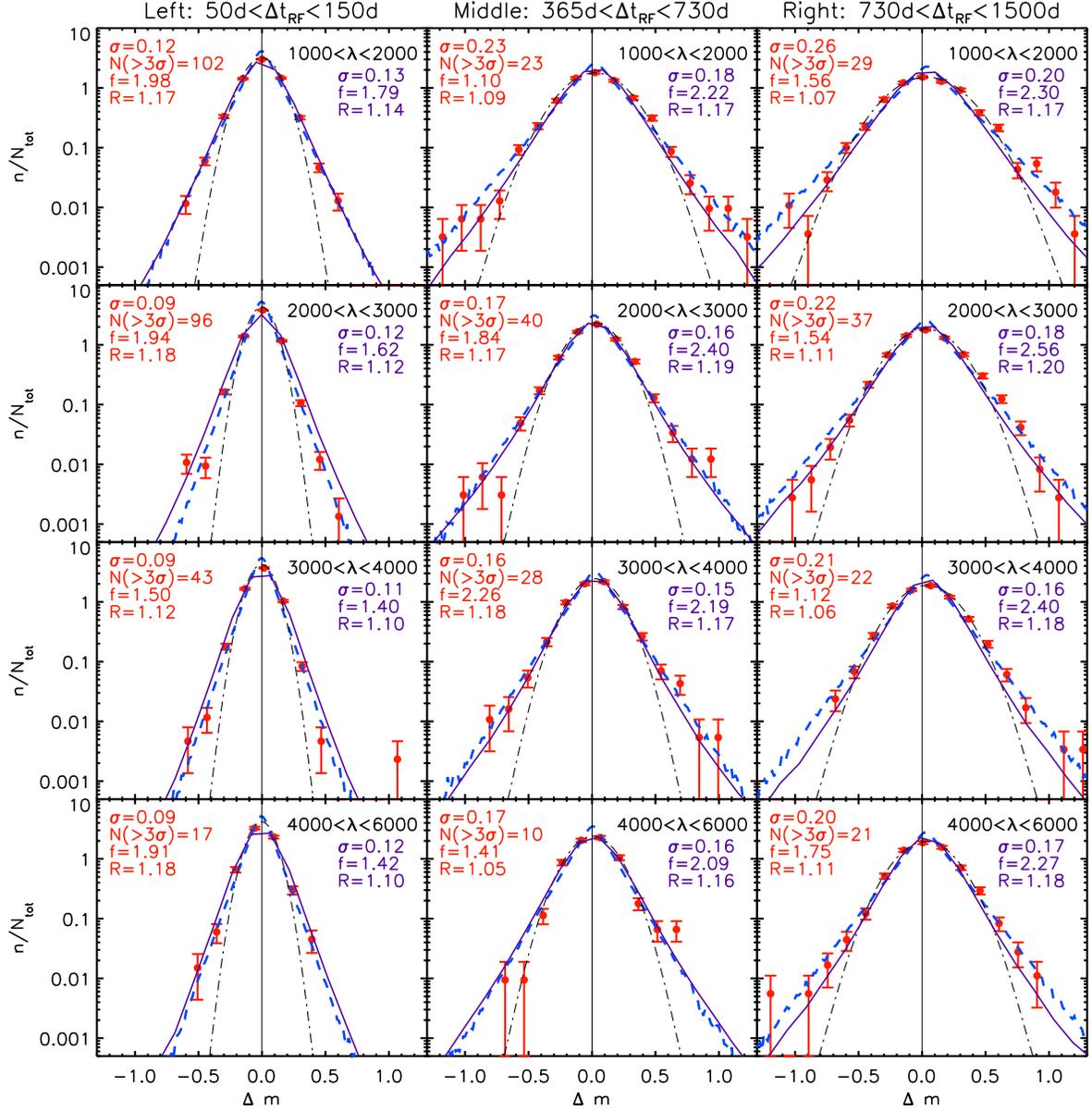}
\caption{\footnotesize As in Fig.~\ref{dmHistObs}, except that the symbols show the
  distribution of measured magnitude differences, $\Delta m$, for
  subsamples selected using rest-frame quantities, as marked
  in the panels. Each row corresponds to the same wavelength range, and
  each column to the same $\dt_{RF}$ range. The solid lines show the
  predicted DRW model distributions (see Section~\ref{model}) with
  widths $\sigma$ listed to the right of the histograms.  
\label{dmHistRestFrame}}
\end{figure*}

Figure~\ref{dmHistRestFrame} shows the distribution of \dm\ in four
bins of rest wavelength and three bins of rest-frame time lag.  The
distributions remain exponential at large $|\dm|$ for all 12 combinations
of $\dt_{RF}$ and $\lambda_{RF}$, as illustrated by the dashed lines.   
The model distributions, which carry information about each individual 
quasar's expected DRW parameters (and thus incorporate the $M_i$ and
$M_{BH}$ information), are shown as solid curves.  The observed and predicted 
SF$(\Delta t_{RF})_{qso}$ values ($\sigma$) are listed to the right and left
of each histogram, respectively. The model and data distributions
agree well, showing that \emph{the exponential distributions seen in
  the statistics of ensembles of quasars naturally result from
  averaging over quasars that are individually well-described by a
  Gaussian DRW process.}  
There are two systematic discrepancies, however. 
First, while the exponential tails can be reproduced at large rest-frame
time lags ($500<\Delta t_{RF}<1500$ days), the value of the observed SF is
systematically higher. Since there are no known intrinsic differences
in the physical properties of the S82 quasars and the two-epoch sample
studied here, the discrepancy is likely due to biases in estimates of
the DRW variability parameters for the S82 quasars.  Secondly, at small 
time lags ($50<\Delta t_{RF}<150$ days), the model over-predicts the data 
rms by 10\%, while the statistical uncertainty is 1\% assuming a perfect 
Gaussian distribution of \dm\ and no other systematic errors. This 
discrepancy may be due to some systematic effect that was not
accounted for when computing the error bars (or when correcting forthe
SDSS photometric errors).  Alternatively, the discrepancy could 
result if the DRW model is inaccurate on these time scales (1--200
days). These discrepancies are discussed further in
Section~\ref{compare}. 

\subsubsection{The SF as a Function of Time Lag and Wavelength}

\begin{figure}[t!]
\epsscale{1}
\plotone{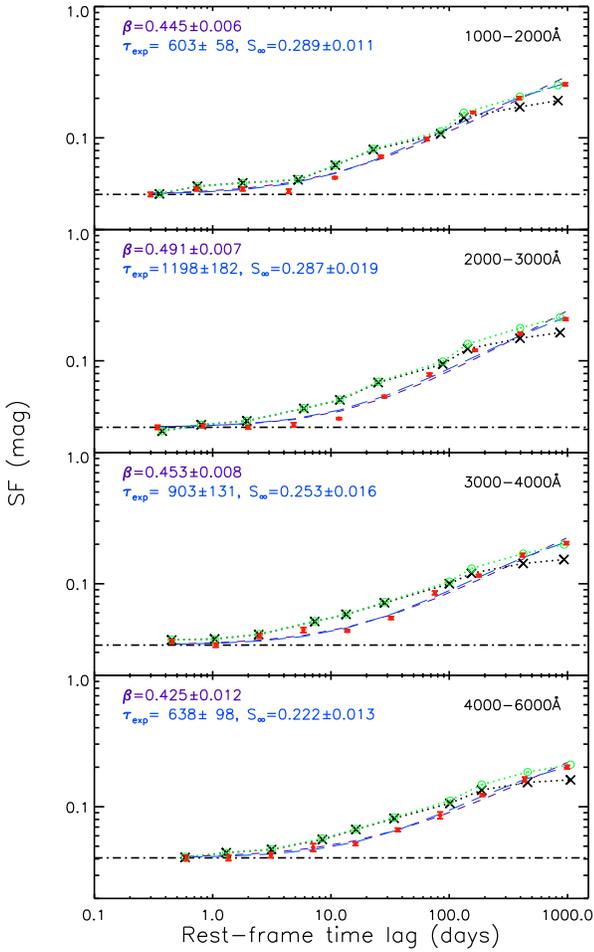}
\caption{\footnotesize The red symbols show the structure function for
  quasar variability as a function of rest-frame time lag, for four
  ranges of rest-frame wavelength, as marked. The short-dashed line in
  each panel is the best-fit power law (${\rm SF} \propto\dt^\beta$),
  with $\beta$ marked in each panel. The long-dashed lines are the best-fit  
  exponential curves, ${\rm SF} = S_{\infty}[1-\exp(-\dt/\tau_{\rm
  SF})]^{1/2}$, with the parameters listed in each panel. 
  Both curves include a photometric noise component equal to the minimum 
  measured SF in each panel (marked by the horizontal dot-dashed line). 
  The dotted lines with cross symbols show the DRW model prediction
  (for 100 realizations, see Section~\ref{model}).  The green dotted
  lines with circles show the prediction of the DRW model trained on
  S82 data with all $\tau$ and SF$_{\infty}$ values multiplied by 2
  and $\sqrt{2}$, respectively. 
\label{SFdt}}
\end{figure}

\begin{figure}[t!]
\plotone{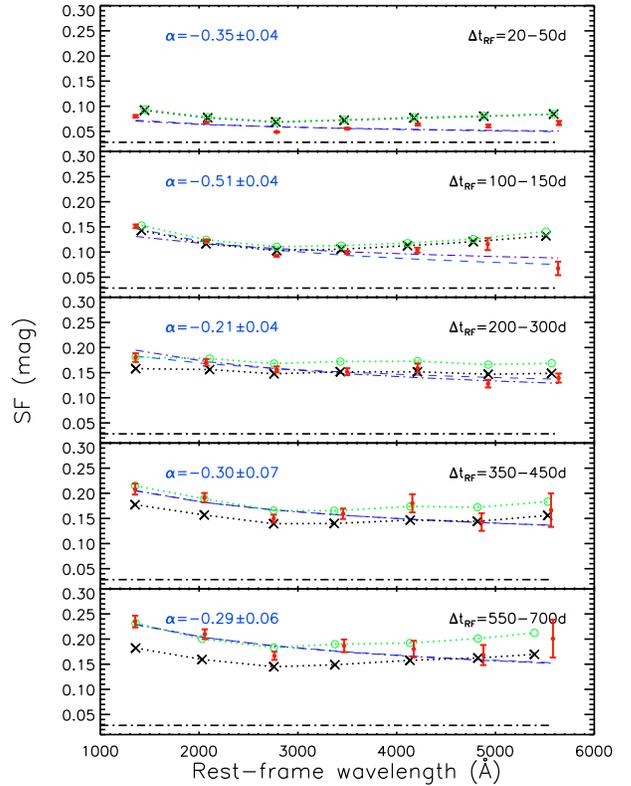}
\caption{\footnotesize The red symbols show the structure function for quasar
  variability as a function of rest-frame wavelength, for five ranges
  of rest-frame time lag, as marked.  The dashed lines are the
  best-fit power laws (${\rm SF} \propto\lambda^{\alpha}$) with the
  power-law index listed in the top-left corners. The dash-dotted
  lines show the functional dependence adopted by I04
  ($\alpha=-0.3$). Both curves include a photometric noise component
  corresponding to $\sigma_{phot}=0.02$~mag. The dotted lines with
  cross symbols show the DRW model prediction (for 100 realizations,
  see Section~\ref{model}).  The green dotted lines with circles
  show the prediction of the DRW model trained on S82 data with all
  $\tau$ and SF$_{\infty}$ values multiplied by 2 and $\sqrt{2}$,
  respectively.
\label{SFw}} 
\end{figure}

Now that we understand and can reproduce the shape of the distribution
of magnitude changes \dm\ between two times, we can simply consider
the SF (i.e., the width of this \dm\ distribution) as a function of
physical quantities such as $\dt_{RF}$ and $\lambda_{RF}$ and test the
model prediction. Figure~\ref{SFdt} shows the SF as a function of
$\dt_{RF}$. Here, the measured SF is not corrected for the SDSS photometric
accuracy. Again, it is apparent that the model predicts a
systematically lower SF for time lags less than 200 days. However, the
stochastic model predicts a systematically higher SF for the longest
time lags.  This suggests the model SF may be biased low at long time
lags, and the bias seems to be most prominent at shorter wavelengths
(Figure~\ref{SFw}).  Note 

\clearpage

\clearpage

\begin{figure}[t!]
\plottwo{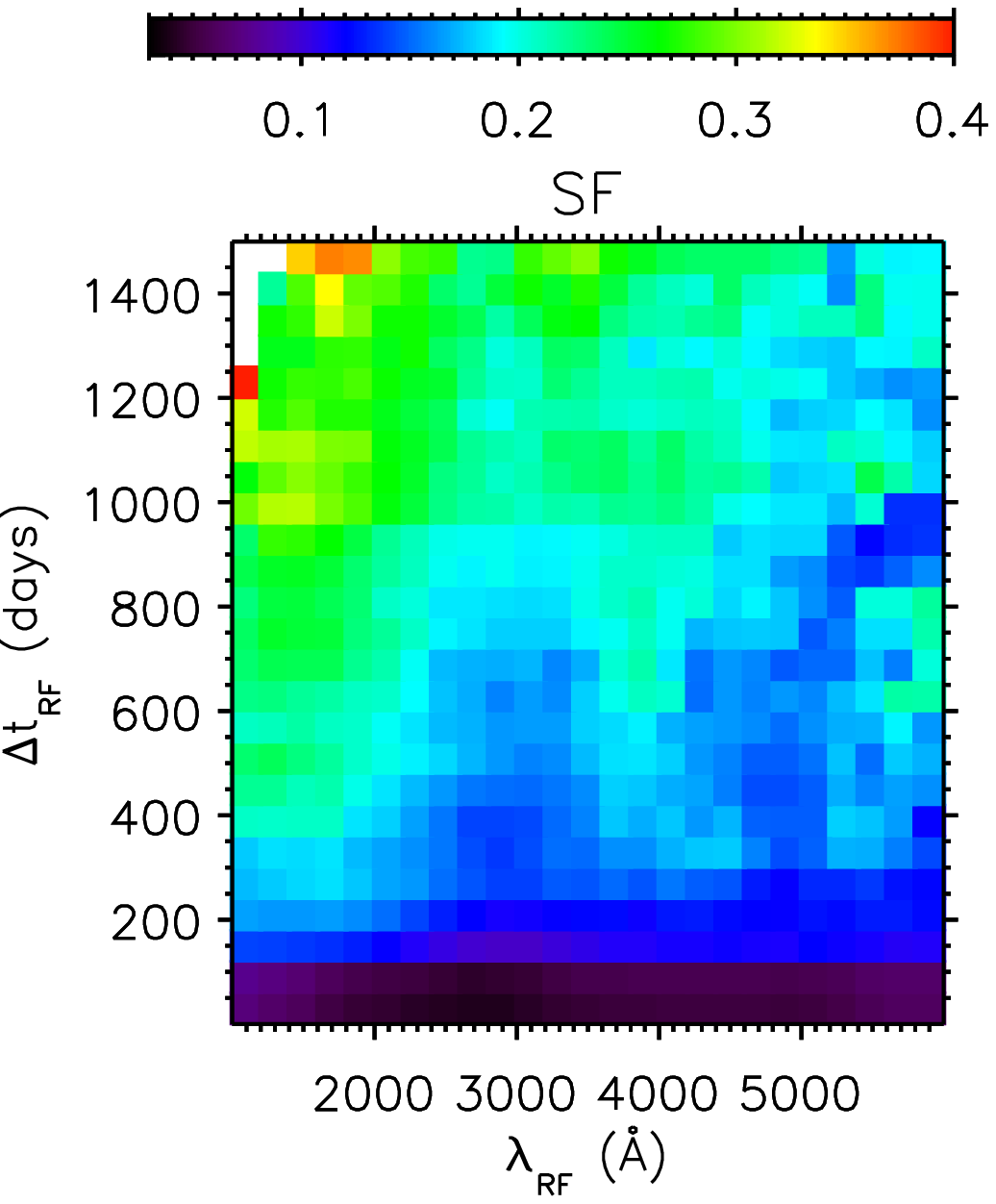}{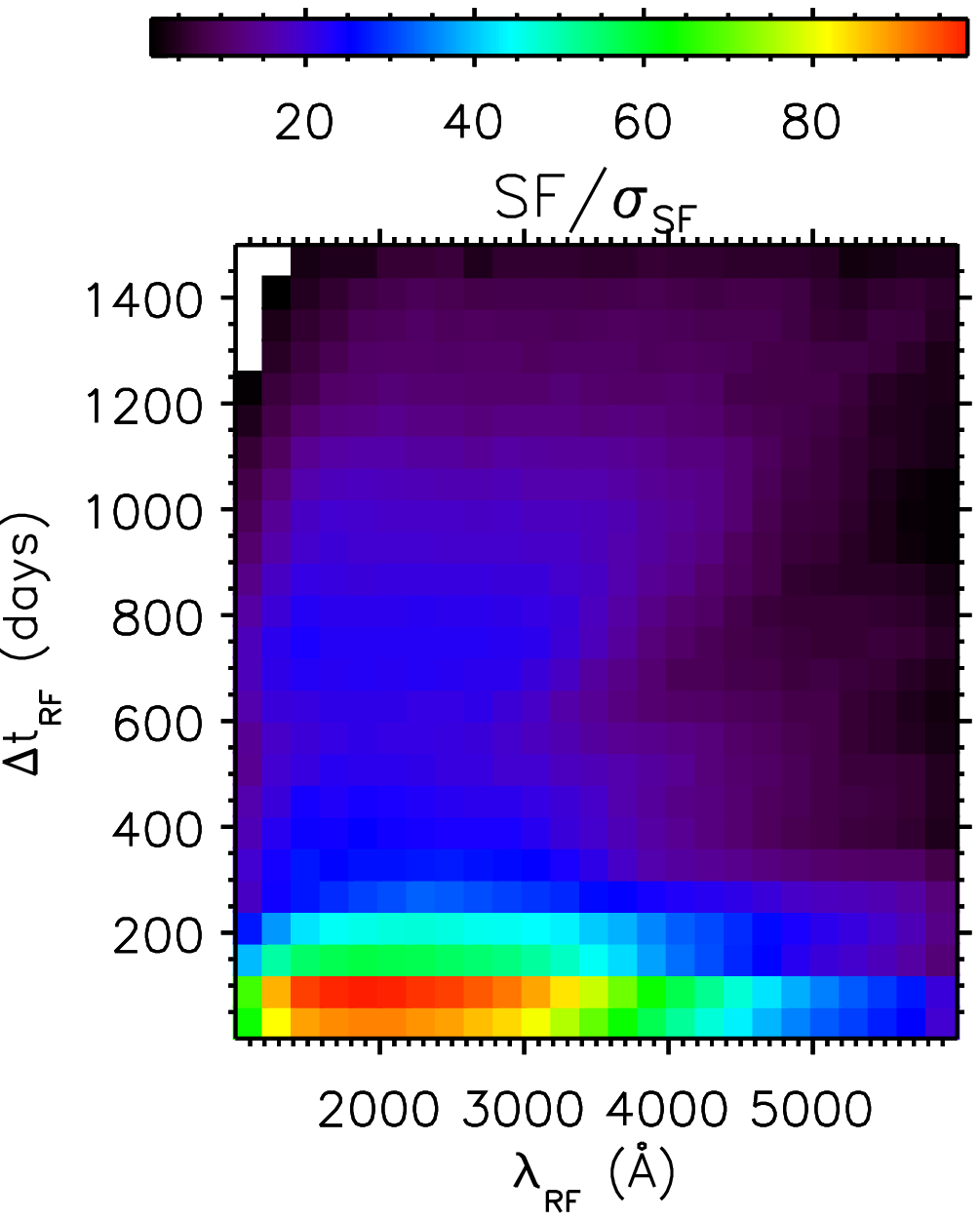}
\plottwo{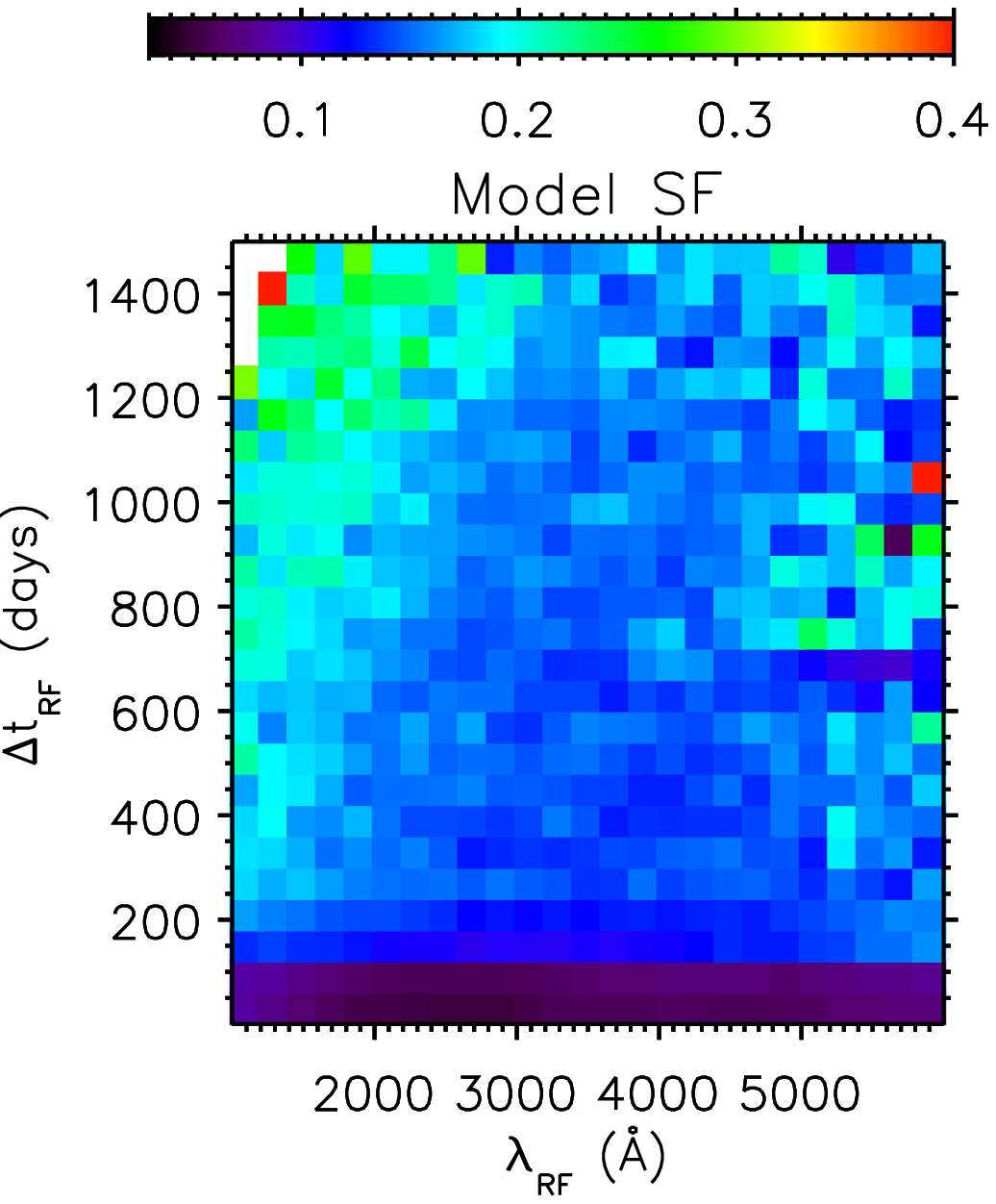}{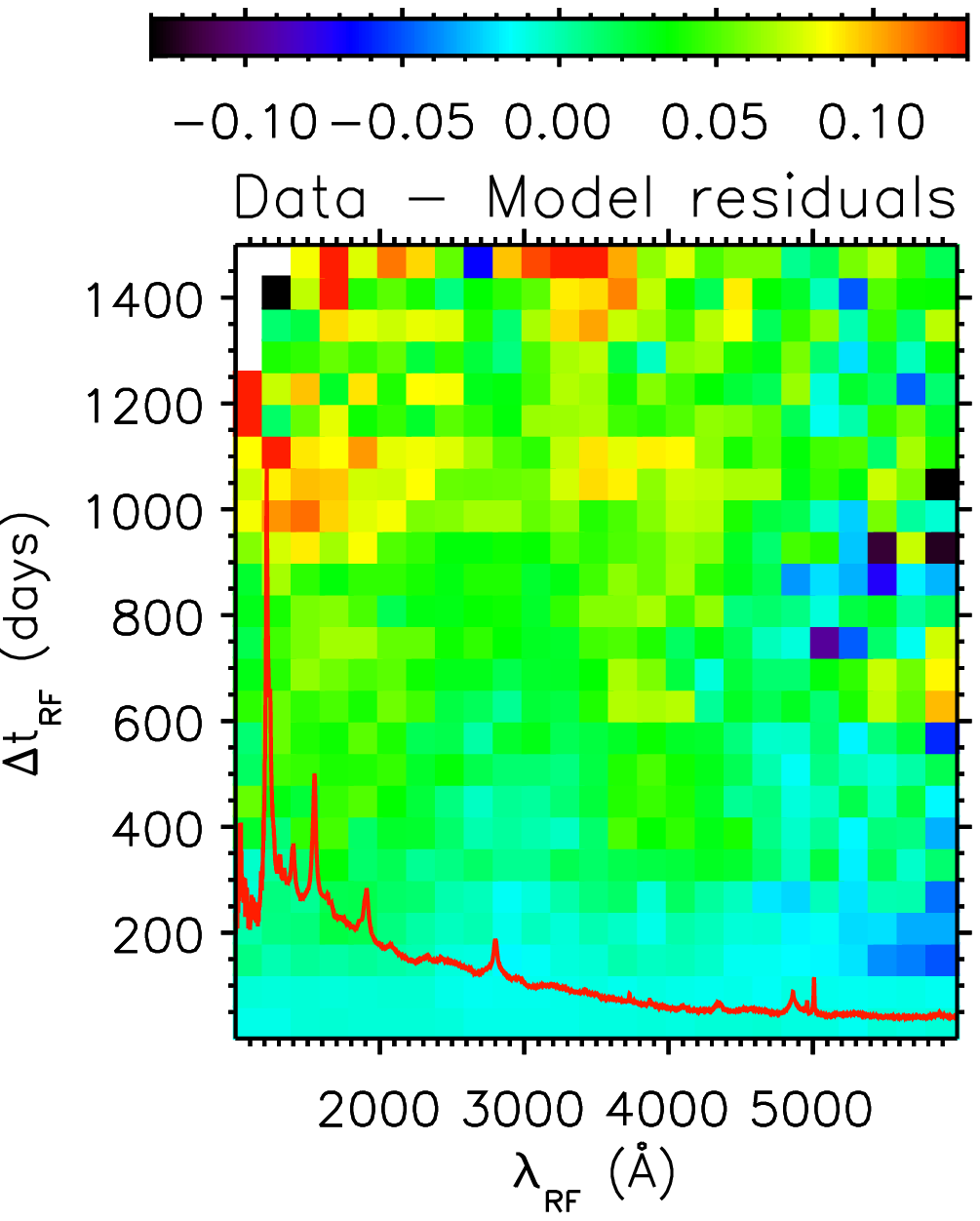}
\caption{\footnotesize The top-left panel displays the structure
  function for quasar variability measured by the SDSS as a function
  of rest-frame wavelength and time lag. SF is computed as
  0.74(IQR)$/\sqrt{N-1}$, where IQR is the 25\% -- 75\% interquartile
  range. The figure is based on 79,787 SDSS measurements, and shows
  the structure function on a linear scale from 0.03 to 0.4 mag, as
  indicated in the top bar. The top-right panel shows the SF
  signal-to-noise, where the error $\sigma_{SF}$ is computed as $1.15 SF/\sqrt{N-1}$.  
  The bottom-left panel shows the model structure function (see text),
  and the bottom-right panel shows the (data--model) residuals. The
  red line shows the SDSS composite quasar spectrum in arbitrary flux
  ($F_\lambda$) units from Vanden Berk et al.\ (2001). The weak
  vertical feature in the residual map at $\sim$2800\AA\ is coincident
  with the MgII line visible in the composite spectrum. The white
  pixels in each panel contain fewer than 5 data points even when
  including all adjacent bins (and thus are not used in our analysis). 
\label{CellMedianPanel} \label{CellMedianPanel2}}
\end{figure}

\begin{figure}[t!]
\plotone{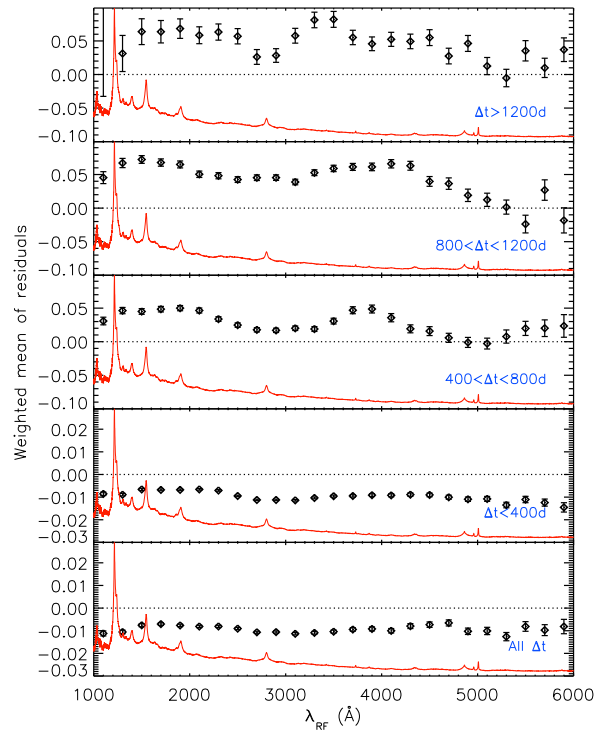}
\caption{\footnotesize Weighted mean of the data--model residuals as a
  functions of rest wavelength (the bottom-right panel of
  Fig.~\ref{CellMedianPanel} collapsed onto the wavelength axis). The
  weighted mean of all time lag bins $i$ is defined as $\sum (SF_i -
  SF_{model,i})w_i/\sum w_i$, where $w_i=1/\sigma_{SF}^2$, the 
  uncertainty from Fig.~\ref{CellMedianPanel}. The error bars of the
  weighted means are computed as $(\sum w_i)^{-1/2}$.
  The composite quasar spectrum is over-plotted in arbitrary flux
  ($F_\lambda$) units.  The dip at 2800\AA\ results from 
  the lagged and smeared variability of the MgII emission line with
  respect to the continuum, and is widened by the errors in estimating
  the rest wavelength from the SDSS bands.  
\label{fig:resdip}}
\end{figure}

\noindent that the measurements at long rest-frame
time lags and short rest wavelengths are dominated by the $u$ band, so
there may be biases simply from the fact that the 14\% of $u$ and $z$
light curves which were dominated by noise were omitted from the
analysis in Mac10. We investigate such biases further in 
Section~\ref{compare}.   
 
In Figure~\ref{CellMedianPanel}, we show the observed ensemble SF
without correction for the SDSS photometric errors. We also show
the signal-to-noise ratio for each bin (top-right panel).  In the
bottom-left panel, we show the model expectation including the
estimated photometric errors. The bottom-right panel shows the
(data~--~model) residuals. Overall, the results are excellent, with a
median difference of only 0.02~mag and a median absolute difference of
0.03~mag. We see a slight decrement at $\dt<200$~days in the residuals
at wavelengths coincident with the 2800\AA\ MgII emission line.  This
decrement is expected based on the results of Reichert et al.\ (1994),
where the MgII emission line is less variable and lags the continuum
fluctuations by almost 10 days.  In Figure~\ref{fig:resdip},
the time axis is collapsed onto the wavelength axis (using the
weighted mean) so that the dip in the residuals is emphasized. 
Due to the excellent SDSS photometry and large sample size, we are
able to resolve this feature. Note that the large width of the
decrement may be due to the fact that the rest wavelengths are
approximated from the fixed effective wavelengths of the SDSS bands.
Not accounting for the variation of the effective wavelength with the
shape and redshift of the quasar spectrum sampled limits the accuracy 
of the approximated rest wavelengths. Thus, any line feature will be
weakened and smoothed by the broad bandpasses of the SDSS 
filters. 
 
\section{Combining Short- and Long-term Quasar Variability
  Measurements (SDSS--POSS)  }
\label{LT}

\begin{figure}[t!]
\plotone{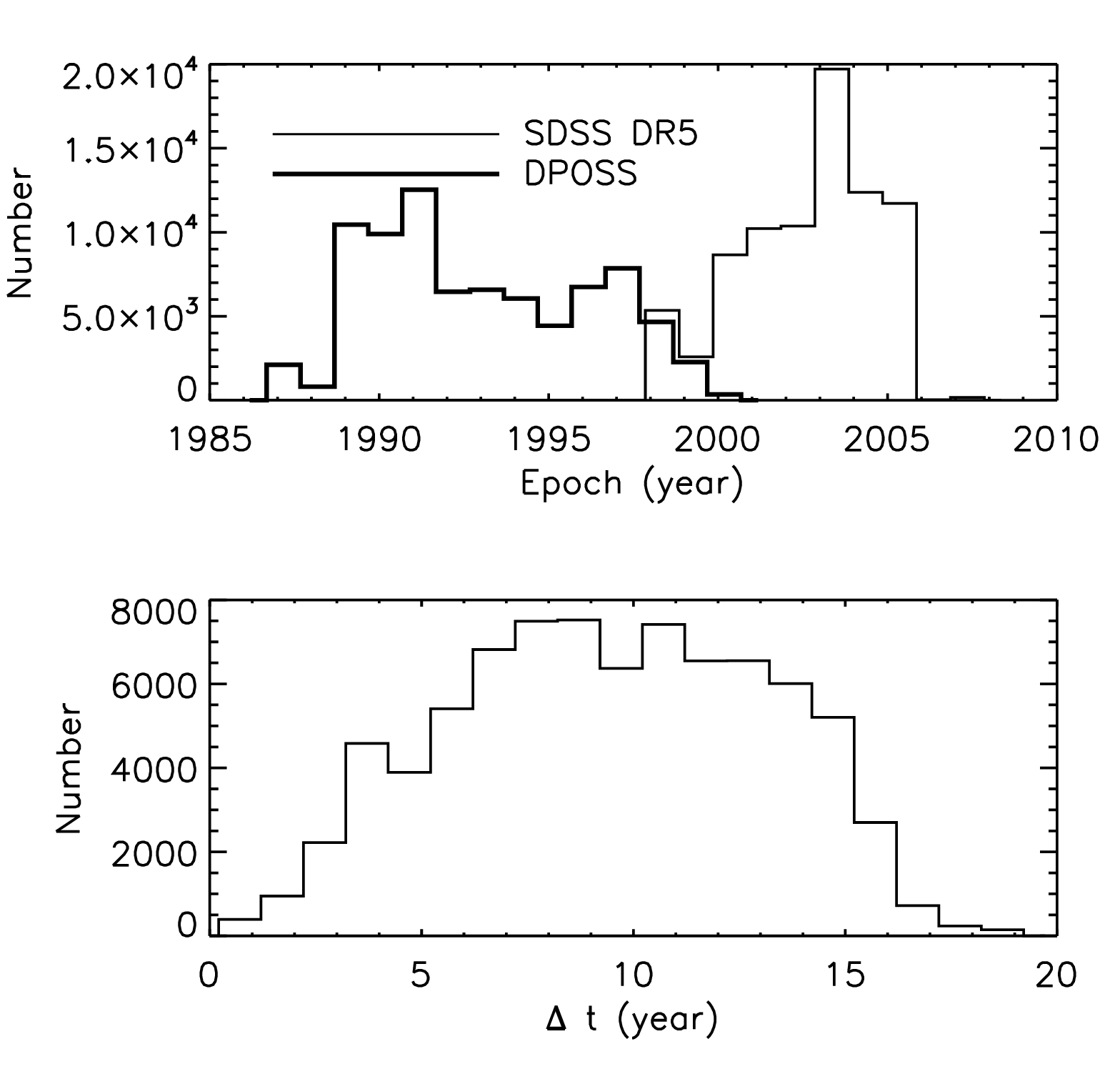}
\plotone{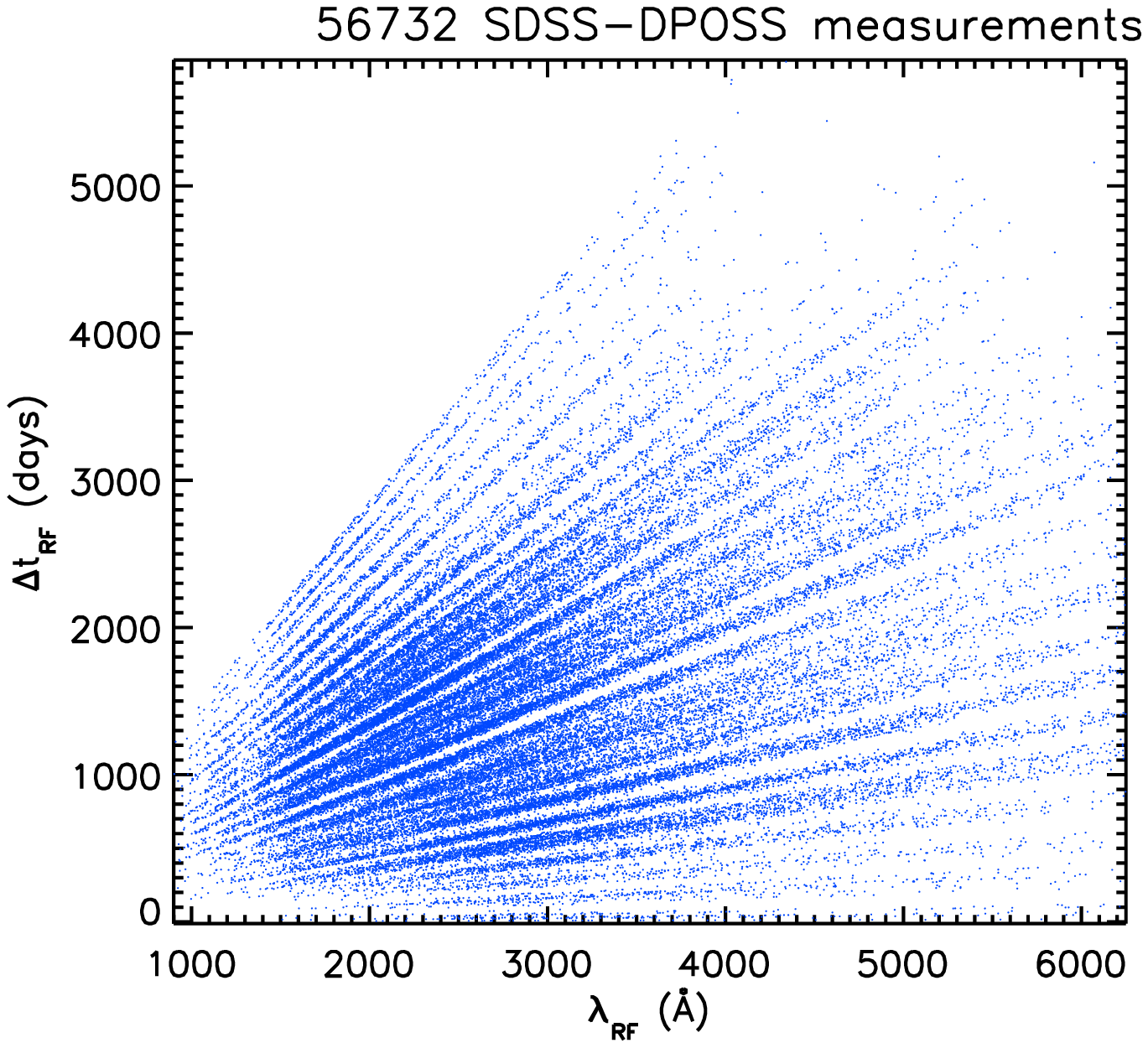}
\caption{\footnotesize Top: Distribution of epochs for the SDSS and
  DPOSS observations.   
Middle: Distribution of the SDSS--DPOSS time lags in the observer's frame.
Bottom: Distribution of the SDSS--DPOSS time lags and wavelengths in the rest frame.
\label{fig:epochs}}
\end{figure}

The exponential distribution analysis in Section~\ref{RF} is sensitive  
to both SF$_{\infty}$ and $\tau$ through the better determined
combination defined by $\hat{\sigma}={\rm
  SF}_{\infty}/\sqrt{\tau}$. However, for the long-term SDSS--POSS data,
the \dm\ distribution is mainly sensitive to SF$_{\infty}$ or $\tau$
at fixed $\hat{\sigma}$. It is then of interest to see whether the
\dm\ distribution remains exponential at large $|\dm|$ and if we can
reproduce the ensemble SF(\dt) on these long time scales. Therefore,
we repeat the analysis in Section~\ref{RF} using the 81,189 SDSS
quasars that are also observed in DPOSS. We only use the DPOSS data for the
\dm\ analysis because the errors for POSS-I are significantly larger.
After applying the data quality cuts, there are 56,732 total \dm\
measurements in the $G$, $R$, and $I$ bands. Figure~\ref{fig:epochs}
shows the distribution of time lags in the observed and rest frames. 

\begin{figure*}
\epsscale{2}
\plotone{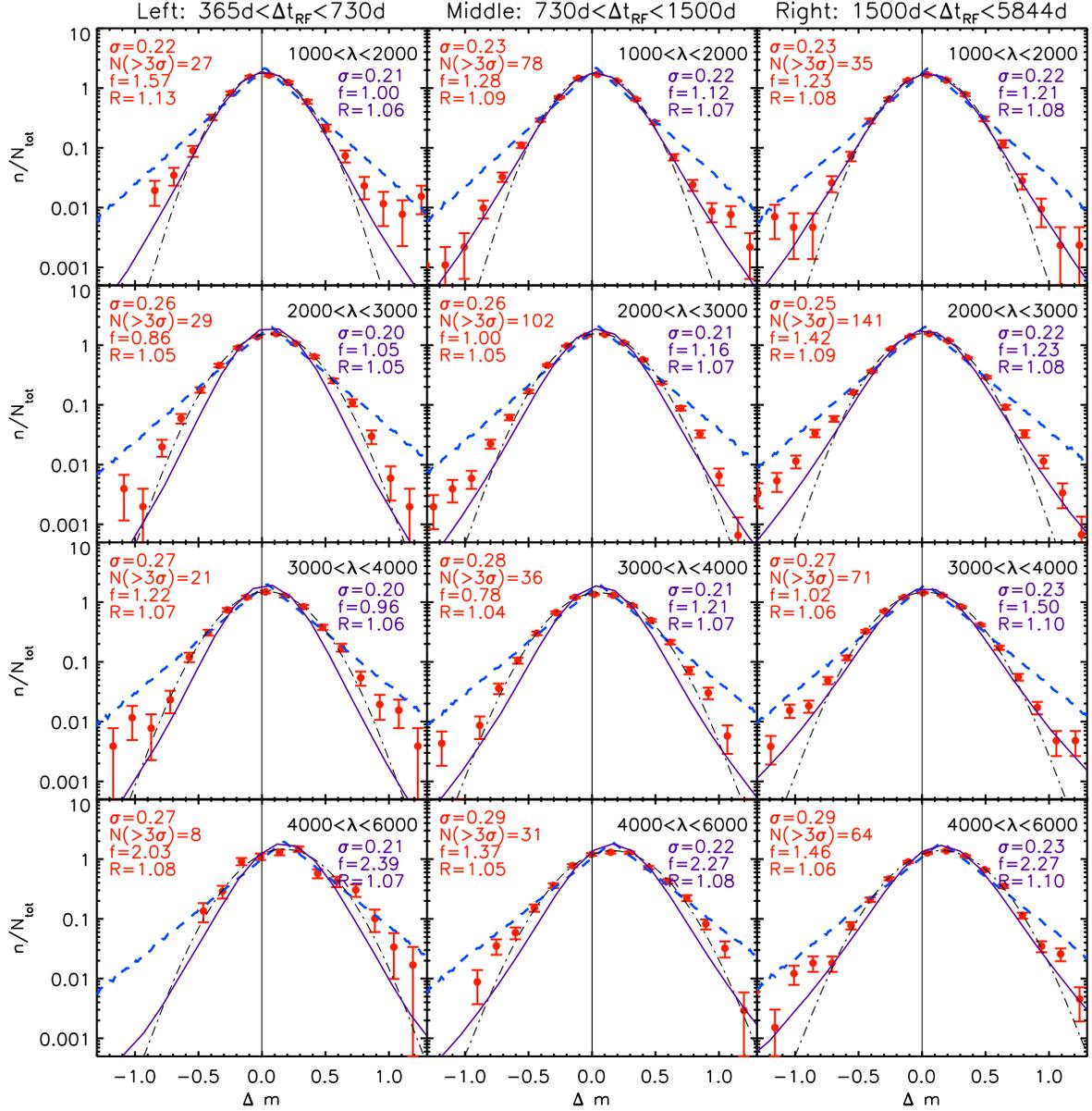}
\caption{\footnotesize Similar to Fig.~\ref{dmHistRestFrame} but for SDSS--DPOSS
  observations over the northern DR5 footprint (with $i<19.1$ in SDSS).
  The photometric accuracy was 
  assumed to be $\sigma_{phot}=0.1$~mag when computing the models. The
  Malmquist-like bias seen in the bottom panels is taken into account by
  shifting the model \dm\ distribution to the right by the observed
  mean. 
\label{fig:dmHistRestFrame_POSS}}
\end{figure*}

The SDSS--DPOSS \dm\ distributions are shown in
Figure~\ref{fig:dmHistRestFrame_POSS}.  Due to the larger photometric 
errors, these distributions appear more Gaussian than those restricted 
to SDSS observations.  In some panels, the peak is also offset from zero towards
positive \dm\ values. This is likely due to the Malmquist-like bias
discussed in dV05, where many objects that were fainter at the time of
the DPOSS observation were then lost and not included. In general, 
the model curves (solid lines) are able to reproduce the shape of the
distributions but underestimate the rms, even when accounting for the
estimated photometric errors of $\sigma_{phot}=0.1$~mag.  

\begin{figure*}
\plotone{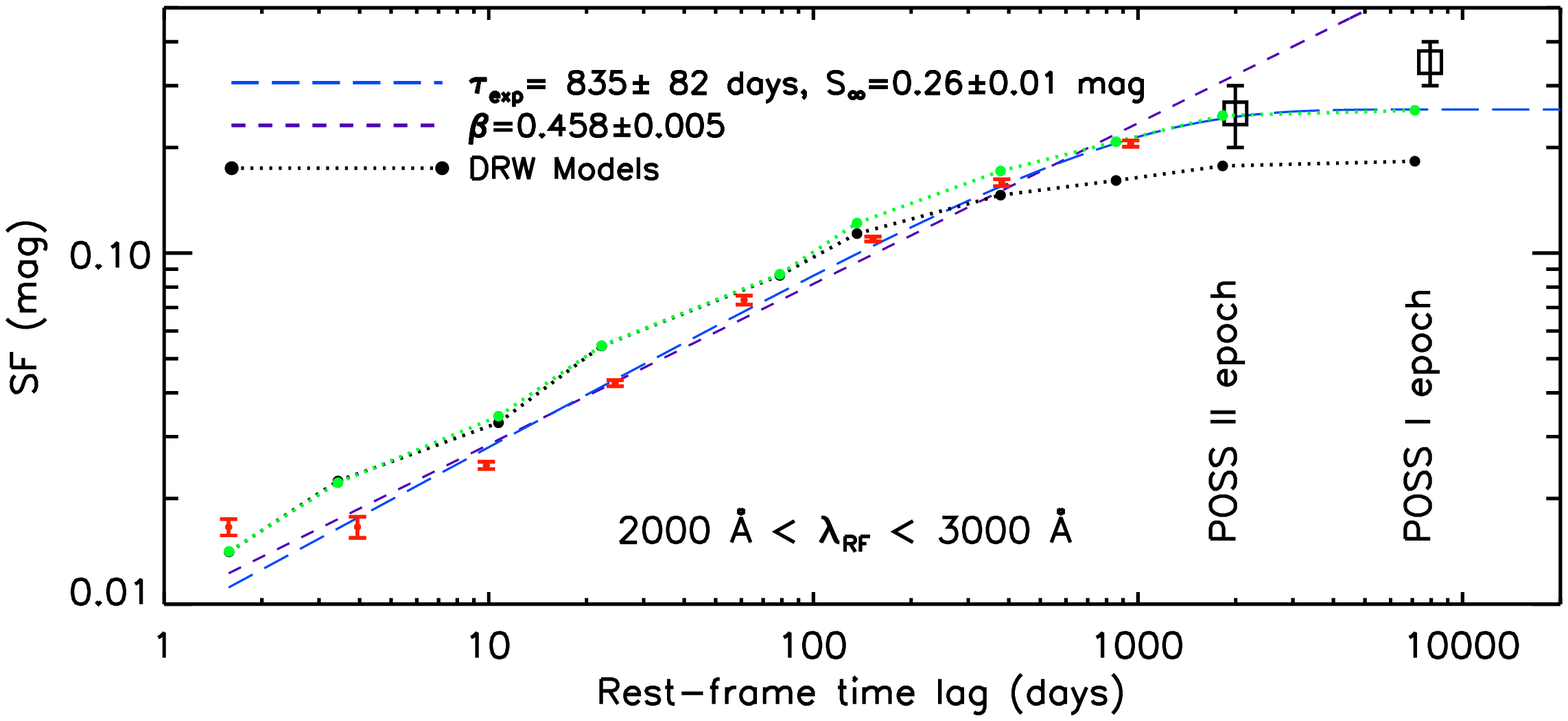}
\plotone{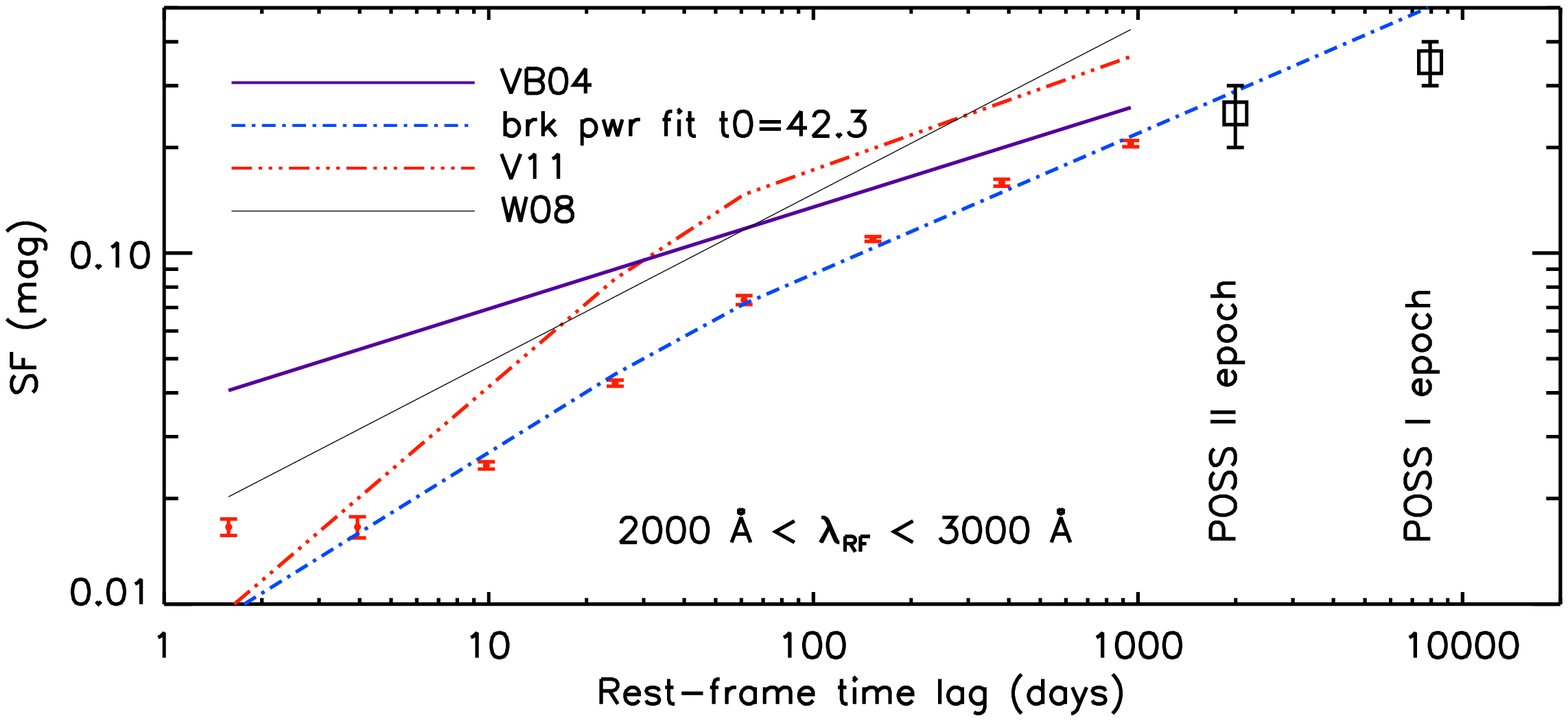} 
\caption{\footnotesize Top: Similar to Fig.~\ref{SFposs}, but plotted against
  rest-frame time lag, for data with rest-frame wavelength in the
  range 2000--3000\AA.  The data have been corrected for photometric
  errors by subtracting 0.018$\sqrt{2}$ in quadrature. The 
  green dotted line shows the prediction of the DRW model trained on
  S82 data with all $\tau$ and SF$_{\infty}$ values multiplied by 2 and
  $\sqrt{2}$, respectively. Bottom: The data are shown along with 
  comparison $g$-band fits from VB04 (purple line), Wilhite et al.\ (2008; thin
  line) and Voevodkin (2011; red dash-triple-dotted line).  The
  normalization for these curves is arbitrary, since we are simply
  comparing the slopes.  The blue dash-dotted line shows the best-fit
  broken power law to our data with the break set to $42.3$~days, as
  in Voevodkin (2011). 
\label{SFall}}
\end{figure*}

Finally, Figure~\ref{SFall} shows the SF over a large range of
rest-frame time lag, including the SDSS--POSS data as measured in
Ses06. We find an overall characteristic time scale of $835\pm 82$~days
with ${\rm SF}_{\infty} = 0.26\pm 0.01$~mag.   Note that in dV05, the
SF reaches $0.46 \pm 0.02$~mag at 40 years in the rest frame 
for the $g$ and $r$ bands combined.  

\subsection{Data--Model Comparison}
\label{compare}
In Figure~\ref{SFall}, we show our model SF as described in
Section~\ref{model}.  We again see that the model is biased to higher
values of the SF between 1 and 200~days.  We attribute this
discrepancy to one or both of the following:  
\begin{enumerate}
\item At short \dt, the errors in our measured SF are underestimated,
  and some source of systematic error, such as an inaccurate
  correction for SDSS photometric errors, causes the measured SF to
  appear smaller than what the model predicts.   
\item The DRW model is inaccurate at short \dt\ and therefore
  overpredicts the ensemble SF. However, this is unlikely given that
  the results using Stripe 82 light curves as well as densely-sampled
  OGLE light curves (Zu et al.~2012) show the DRW model to be a good
  fit on time scales between \about 5 and 200 days.  
\end{enumerate}
Nevertheless, our measured slope of the SF seems well-constrained.  We
again see that beyond 200~days, the model is biased to lower values of
the SF.  The difficulty to exactly reproduce the long-term variability
amplitudes suggests that the distributions of long-term amplitudes
and/or time scales used in our model underestimate the true
distributions.  This bias may result from the following:   
\begin{enumerate}
\item A bias in the measured $\tau$ (and thus SF$_{\infty}$) values 
  for S82 quasars due to insufficient light curve lengths.  
  Since $SF(\Delta t << \tau) = \hat{\sigma}\sqrt{|\Delta t|}$, 
  $\hat{\sigma}={\rm SF}_{\infty}/\sqrt{\tau}$ is the most strongly
  constrained quantity on time scales shorter than $\tau$. That is, on
  a grid of $\log(\tau)$ versus $\log(\rm{SF}_{\infty})$, the best-fit
  values will be scattered due to fitting errors along lines of
  constant $\hat{\sigma}$, but much less perpendicular to it (i.e.,
  along lines of constant $K=\tau\sqrt{\rm{SF}_{\infty}}$). Therefore,
  as the light curve length decreases, the mean best-fit
  $\hat{\sigma}$ will not vary significantly, while the best-fit
  parameters $\tau$, $\rm{SF}_{\infty}$, and $K$ will become biased
  (see Koz{\l}10; Mac10; MacLeod et al.~2011). 
\item A bias in the deterministic model (Eq.~\ref{eq:form}). Due to
  the finite length of the S82 survey, 22\% of the S82 sample was
  excluded from the analysis in Mac10 due to indeterminately long
  time scales. In this case, the measured parameters $\tau$ and  
  SF$_{\infty}$ are accurate for the remaining 78\% of the S82 sample,
  but Eq.~\ref{eq:form} is only accurate for the lower values of
  $\tau$ ($\tau\lesssim10^3$) and SF$_{\infty}$ from which the
  correlations are derived.    
\end{enumerate}
In principle, both effects will lead to lower overall $\tau$ and
SF$_{\infty}$ in the distributions for S82 quasars, which will in turn
cause the model ensemble SF to flatten at $\dt_{RF}<700$~days as seen
here. Indeed, the excluded 22\% of the sample will mostly contribute
power to the long-term rather than the short-term SF given their
indeterminately long time scales. However, when simply including an
additional population of long-$\tau$ objects to fill in the missing
22\%, we are unable to quantitatively reproduce the last SDSS data
point at $\dt_{RF}=1000$~days unless all the additional objects have
$\tau\gtrsim 10,000$~days in the rest frame (while keeping the same  
distribution of $\hat{\sigma}$ as observed). On the other hand, if we
double all the $\tau$ values in our model (instead of adding an
extra population of quasars at long $\tau$), while keeping
$\hat{\sigma}$ fixed so that the SF$_{\infty}$ values are multiplied
by $\sqrt{2}$, we are able to reproduce the observed long-term
SF, as shown by the agreement between the green lines in
Figures~\ref{SFall} and \ref{SFdt} to the data points at long
$\dt_{RF}$.  In summary, the best-fit $A$
coefficients from Eq.~\ref{eq:form} need to be altered upwards by 0.15
dex in the case of SF$_{\infty}$, and by 0.30 dex in the case of
$\tau$, in order to explain the long time scale constraints provided by
the SDSS--POSS dataset. 

At least half of these corrections can be understood as due to a
fitting bias towards shorter $\tau$ (see Figure~7 in Mac10 and
Figure~15 in MacLeod et al.\ 2011).  Another effect could be due to
uncertain behavior for long time scales. The DRW process corresponds
to a power spectral distribution (PSD) proportional to
$1/f^2$ at frequencies $f>(2\pi\tau)^{-1}$, flattening to a constant
at lower frequencies. Using the S82 data and computational technique
described in Mac10, we were able to rule out an extrapolation of the
$1/f^2$ power-law. However, we were unable to distinguish between a
$1/f^{0}$ or a $1/f$ PSD at frequencies $f<(2\pi\tau)^{-1}$, where the
latter dependence is observed in the X-ray PSDs for Galactic black
holes as well as AGN (McHardy et al.\ 2006; Kelly et al.\ 2011).

We compare the observed SF slopes from Voevodkin (2011), VB04, and
Wilhite et al.\ (2008) to our data in the bottom panel of
Figure~\ref{SFall}. Voevodkin (2011) found that a broken power-law
provides a good fit to the S82 $g$-band ensemble SF with a slope of
0.33, steepening to 0.79 below 42 days.  We also show the best-fit
broken power law to our data with the break fixed to 42 days, and we 
find power law indices at short and long \dt\ of 0.53 and 0.40,
respectively. While our 2-epoch SDSS data are consistent with the
shallower slope of 0.33, our results do not support the conclusion of
a much steeper SF(\dt) for small \dt\ found by Voevodkin (2011) for
either the S82 or 2-epoch data sets.  We interpret the broken power
law form to be a consequence of the turnover in the SF due to
the mean characteristic time scale, as the observed SF is fully
consistent with the form expected for a DRW (Eq.~\ref{eq:sfdt}).

\section{Predictions for Future Surveys}
\label{FS}
Using the observed \dm\ distributions, we can predict the number of
quasars with \dm\ exceeding an arbitrary limit that might be seen in a
survey with a given number of quasars. This information is useful for
transient identification, in particular to identify quasars that
contaminate candidate lists of other objects.  For example, Vanden
Berk et al.\ (2002) reported an orphan gamma-ray burst afterglow based
on a 2.5 magnitude decrease in the optical flux of an unidentified
point source. Instead, as pointed out by
Gal-Yam et al.\ (2002), the observations are best explained by the
presence of a spectroscopically identified, highly variable quasar. 

In order to quantify the importance of quasar contamination in future
transient surveys, such as the Palomar Transient Factory (PTF; Law et
al.\ 2009) and the Large Synoptic Survey Telescope (LSST; Ivezi{\'c}
et al.\ 2008), we need to know the probability that a quasar can
increase brightness by \dm\ (over the faint survey limit) within a
time \dt. For example, assume that no source is detected above a faint
limit of $m_{faint}$ on a given night, but when repeating the
observation some time later, a source is detected with magnitude $m =
m_{faint} - \dm$. In this case we would like to know how many quasars
with a \dm\ at least as large as observed could be present in a
particular scanned area and to a faint limit of $m_{faint}$. Given
that our stochastic model performs well at reproducing the observed
ensemble variability of quasars,  we can make robust and useful
predictions for future surveys. 

We make use of a mock LSST quasar sample which includes absolute $B$
magnitudes ($M_B$) and redshifts generated over 100~deg$^2$ of sky
using the luminosity function from Bongiorno et al.\ (2007) for the
purposes of LSST image simulations.  We limit the sample to $M_B<-20$
following Table 10.2 in the LSST Science Book (LSST Science
Collaborations and LSST Project 2009). The distance modulus is
computed assuming a standard cosmology: $\Omega_m=0.27$,
$\Omega_{\Lambda}=0.73$, and $h=0.71$.  The LSST magnitudes ($m$) and
rest-frame $M_i$ (based on $M_B$) are computed using the composite
quasar spectrum from Vanden Berk et al.\ (2001).  The black hole
masses are estimated from the $M_i$ values using the prescription in
Mac10.   

We generate magnitude differences for each quasar in the mock LSST
sample based on the DRW model most appropriate for the quasar's
physical parameters, as described in Section~\ref{model}. When
computing the DRW model, the $\tau$ (SF$_{\infty}$) values are
multiplied by 2 ($\sqrt{2}$) with respect to the expected values based 
on Mac10 in order to correct for the bias due to limited time sampling
in S82 (see previous section).  We note that 
correcting for this bias should also account for at least
some of the residual scatter in $K$ estimated in Section~\ref{model}; 
however, for simplicity we still include this scatter along with the
bias correction in our simulations.
First, we consider three different survey faint limits of
$m_{faint}=19.1$ (to establish similarity with the SDSS results),
$m_{faint}=22$, and $m_{faint}=24.5$, excluding all mock quasars with
$m\geq m_{faint}$ in each case. For the $m < 19.1$, 22, and 24.5
simulations respectively, 1000, 100 and 11 model \dm\ values are
generated per quasar to increase the sample size. 
A Gaussian noise component of width
$\sqrt{2}\sigma_{phot}$ is added to all curves to simulate a
photometric accuracy similar to the SDSS (the photometric accuracy for 
future surveys such as LSST will likely be better than that for the
SDSS, but this is a higher order question than investigated here).  
Note that we retain all \dm\ values in the simulations, including $|\dm|>3$. 

\begin{figure}[th!]
\epsscale{1}
\plotone{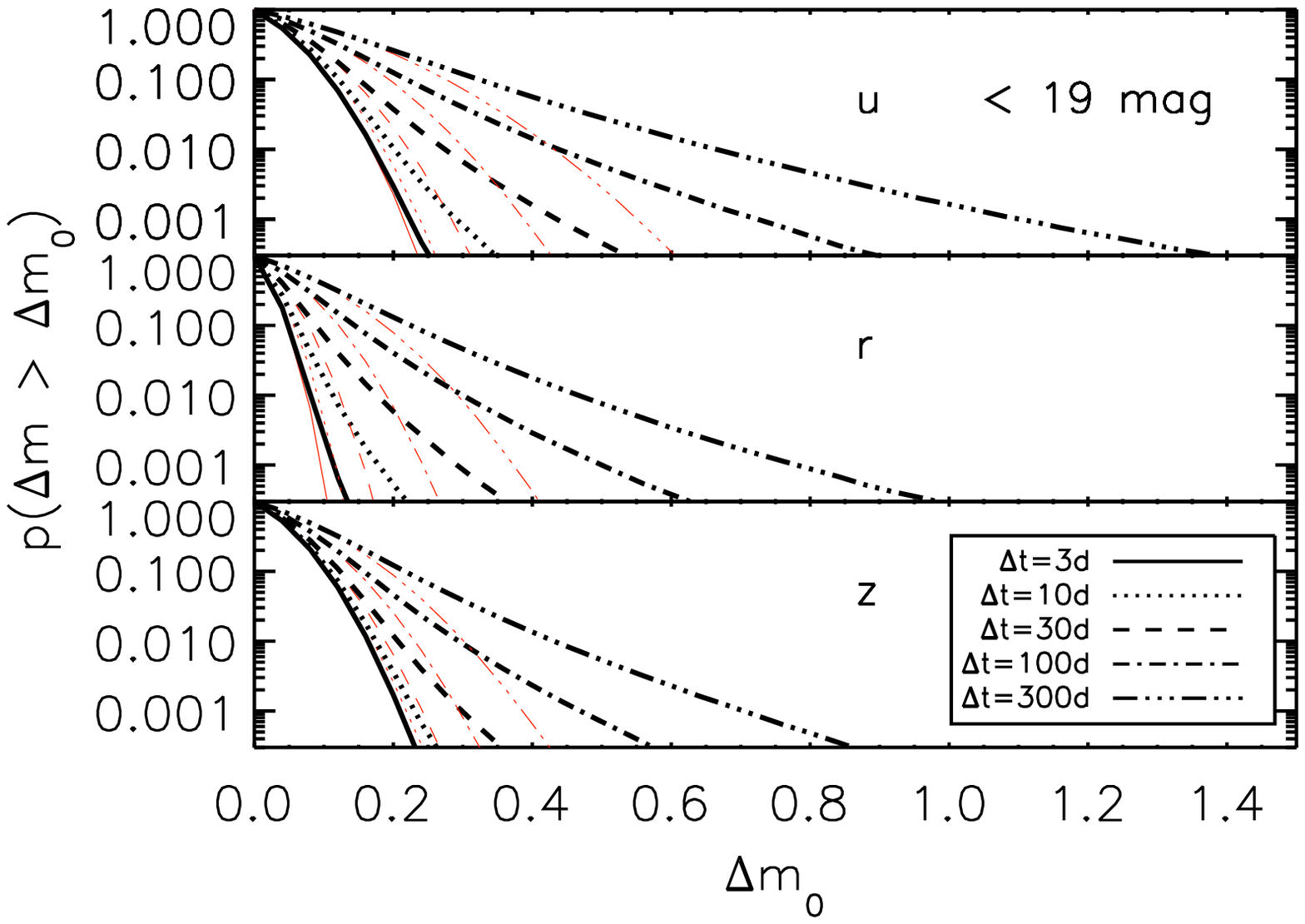}
\plotone{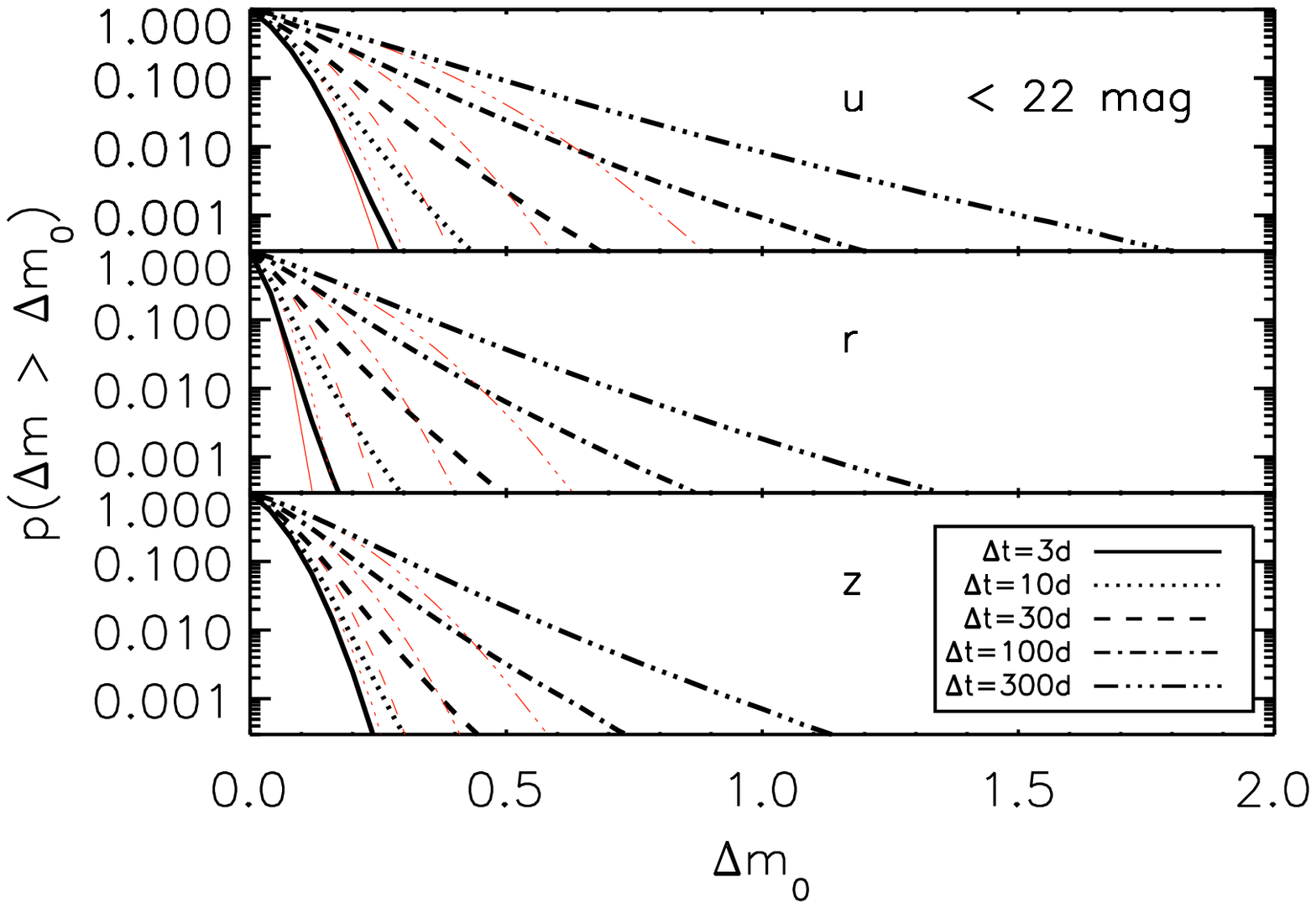}
\plotone{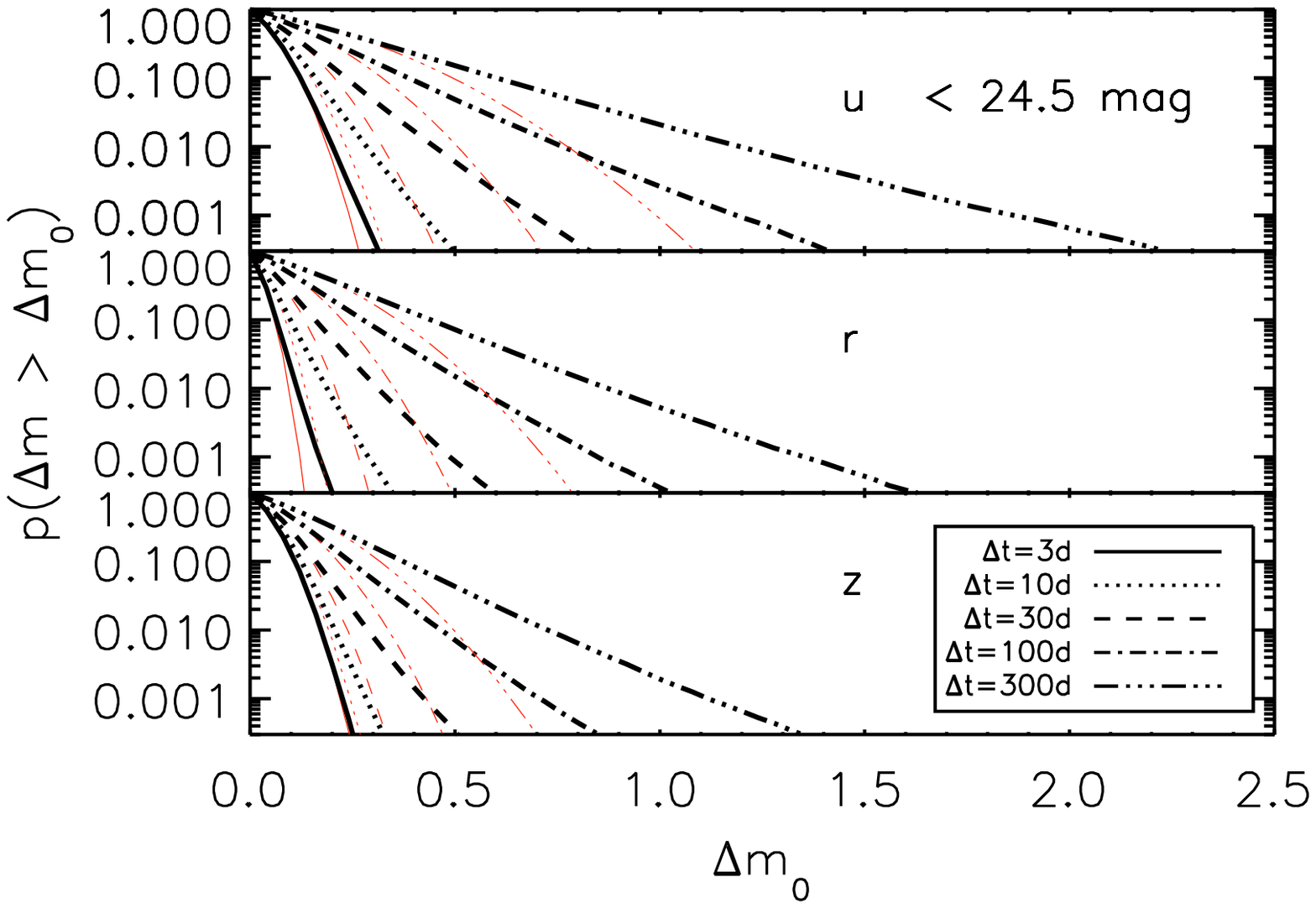}
\caption{\footnotesize The
  predicted cumulative distribution of magnitude differences in $u$,
  $r$, and $z$ bands as a function of observed time lag ($\Delta t$) for \about
  1 million quasars with magnitudes less than 19.1 (top three panels),
  22 (middle three panels), and 24.5 (bottom three panels). 
  The thin curves show the Gaussian distributions with the same rms widths.
\label{dmpredict}}
\end{figure}

Figure~\ref{dmpredict} shows the simulated cumulative distribution of 
\dm\ in the $urz$ bands for three faint magnitude limits. Also shown 
are Gaussian analytic functions as thin curves with the same rms as
the data.  Using these cumulative distributions, which are based on
\about $10^6$ mock quasars, we predict probabilities down to our
resolution limit ($P\gtrsim 10^{-6}$).  Table~\ref{tab:dmpredict}
lists the predicted probabilities of observing a quasar with $m<24.5$
and a magnitude difference of $\dm>1$~mag and $\dm>2$~mag over 3, 30,
and 300 days, using the observed frame, in the $urz$ bands. Due to the
exponential nature of the \dm\ distributions, the probability of
observing $\dm>1$~mag reaches 0.02 in the $u$ band (where variability
is strongest) for time lags of 300~days, and $6\times 10^{-4}$ for
$\dm>2$~mag. Assuming Gaussian \dm\ distributions will result in
erroneous probability estimates of $9\times 10^{-4}$ and $10^{-6}$,
respectively. 

\begin{figure}[th!]
\plotone{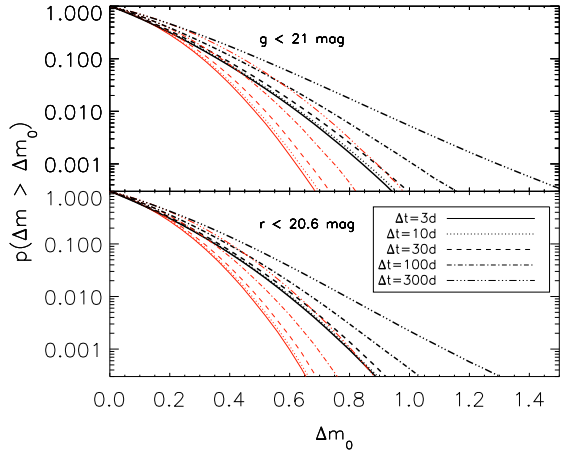}
\caption{\footnotesize The predicted cumulative distribution of
  $g$-band (top) and $r$-band (bottom) magnitude differences,
  including realistic PTF photometric errors as a function of observed
  time lag ($\Delta t$) for quasars brighter than the PTF magnitude
  limit ($g<21.0$, $r<20.6$). The thin curves show the Gaussian
  distributions with the same rms widths.
\label{dmpredictPTF}}
\end{figure}

Next, we adopt realistic PTF photometric errors which vary as a
function of magnitude, and repeat the simulation in $g$ and $r$ using the
PTF magnitude limits ($r<20.6$, $g<21.0$). The errors range from 0.008~mag at
the brightest limits to 0.2~mag at the faint end. 
As shown in Figure~\ref{dmpredictPTF}, the resulting distributions are
more Gaussian than those in the previous simulations due to the
different dependence of errors on magnitude.  However, at extreme values of \dm,
the Gaussian curves significantly underestimate the detection
probabilities in both $g$ and $r$. 

These results would have been useful to Vanden Berk et al.~(2002), who
reported a single transient with $\dm=2.5$~mag over 410~days in the
$gri$ bands in early SDSS data covering $1,500~{\rm deg}^2$. 
Since the object had the spectrum of a normal galaxy 
in its faint phase (rather than a nonstellar spectrum with broad emission lines
indicative of quasars), the authors concluded that it was more likely
to be a gamma-ray burst afterglow than a highly variable quasar, given
the large drop in flux.  
We find that the probability of a quasar having $\dm=2.5$~mag over
410~days with $m_{faint}=22$ is $10^{-5}$ in the $g$ and $r$
bands. To find the number of quasars expected to exhibit this
variability in $1,500~{\rm deg}^2$ of sky, one needs to know the
quasar density. We extrapolate Figure~13 in Richards et al.~(2006) to
a density of \about $120~{\rm deg}^{-2}$ at $i<22$.  Therefore, we expect
roughly $10^5$ quasars in $1,500~{\rm deg}^2$ to $i<22$.
Given a probability of $10^{-5}$, we expect that roughly one quasar will show
$\Delta r=2.5$~mag over $\dt=410$~days in the SDSS sample studied by Vanden
Berk et al.~(2002).  This is consistent with their finding, and with
the follow-up study of Gal-Yam et al.~(2002) who confirmed that it is
indeed a highly variable quasar.  If one
assumes a Gaussian distribution of \dm, the probability is 
less than $10^{-6}$,
and one would expect $\lesssim 0.1$ quasar to be found with these
parameters.  Given this expectation value, the probability to detect
one quasar is at most $0.09$, and the quasar hypothesis can be (erroneously)
rejected at a $\gtrsim 2\sigma$ level. 

\begin{deluxetable}{r c c c}
\tablewidth{0pt}
\tablecaption{Predicted \dm\ Probabilities for 
  $m<24.5$ \label{tab:dmpredict}} 
\tablehead {\dt\ (days) & Band & P($\dm>1$~mag) & P($\dm>2$~mag)}
\startdata
3   & $u$ & $<2\times 10^{-6}$ & $<2\times 10^{-6}$ \\ 
3   & $r$ & $<2\times 10^{-6}$ & $<2\times 10^{-6}$ \\
3   & $z$ & $<2\times 10^{-6}$ & $<2\times 10^{-6}$ \\
30  & $u$ & $7\times 10^{-5}$  & $<2\times 10^{-6}$ \\ 
30  & $r$ & $4\times 10^{-6}$  & $<2\times 10^{-6}$ \\
30  & $z$ & $<2\times 10^{-6}$ & $<2\times 10^{-6}$ \\
300 & $u$ & 0.02              & $6\times 10^{-4}$ \\
300 & $r$ & 0.005             & $6\times 10^{-5}$ \\ 
300 & $z$ & 0.002             & $1\times 10^{-5}$ \\
\enddata
\end{deluxetable}

\section{                       Discussion                       } 
\label{Disc}

We have assembled, organized and publicly released a dataset including
  \about 3.5 million photometric measurements for 80,000
  spectroscopically confirmed quasars.  The available time lags span
  0.8~days to almost 20~years in the observer's frame. We have
  analyzed and quantified the observed variability in the observer's
  and rest frames. By assuming a DRW model for each quasar in our
  sample, we reconcile the observed variability of individual quasars
  in S82 with their ensemble statistics. Our principal results are as
  follows:   
\begin{enumerate}
\item  Long-term quasar variability measurements, constrained using
   SDSS and POSS data for time lags up to 50~years (in the observer's
   frame), conclusively show that a simple power-law dependence for
   the structure function cannot be extrapolated beyond a decade,
   and suggests an average characteristic time scale for quasar
   variability in the rest frame of $\sim$2~years and an average
   long-term dispersion of $\sim$0.26~mag (for rest wavelengths
   2000--3000\AA).  This behavior extrapolates well to the UV results 
   of Welsh, Wheatley \& Neil (2011), who find that the SF for GALEX
   NUV data reaches about 0.4~mag and flattens at $\dt_{RF}>300$~days.
   This SF limit corresponds to a limiting SF value of 0.33~mag when using our
   definition of the SF (see Section~\ref{SF}), and 0.27~mag when also
   scaling to the $u$-band using the wavelength dependence from
   Eq.~\ref{eq:form}. This result is in close agreement with the SF at the
   shortest wavelengths in our data set. Voevodkin (2011) found that a
   broken power-law provides a good fit to the S82 ensemble SF with
   a slope of 0.33 at long time scales steepening to 0.79 below
   42~days.  Our 2-epoch SDSS data are consistent with the
   shallower slope of 0.33, but our data do not support the conclusion of
   a much steeper SF(\dt) for small \dt\ found by Voevodkin
   (2011). While we cannot rule out a broken power-law dependence with
   the available data, the observed SF is fully consistent with the
   form expected for a DRW (Eq.~\ref{eq:sfdt}).      
\item We tested the DRW model results based on SDSS Stripe 82 data on
  an independent dataset, and confirm that the variability parameters
  $\tau$ and SF$_{\infty}$ correlate with physical parameters as found 
  for individual quasars (e.g., Mac10 and references therein).  This
  is evident from the agreement of our model with the observed
  ensemble variability of SDSS quasars.  However, the results indicate
  that the measured $\tau$ and SF$_{\infty}$ distributions are biased
  low for the S82 sample by a factor of about 2 and $\sqrt{2}$,
  respectively. This bias most likely results from the 10-year length
  limit of the S82 light curves, although it could also be due to
  uncertain behavior for long time scales.  The best-fit $A$
  coefficients from Eq.~\ref{eq:form} (Mac10) need to be shifted upwards by 0.15
  dex in the case of SF$_{\infty}$, and by 0.30 dex in the case of
  $\tau$, in order to explain the long time scale constraints provided by
  the SDSS--POSS dataset. These shifts leave the shorter time scale
  variability statistics unchanged. 
\item For a given time lag and wavelength, the magnitude difference (\dm) 
  distribution is exponential rather than Gaussian for large magnitude
  changes. This is well explained as a cumulative effect of
  averaging over quasars with a range of different $\tau$ and ${\rm
  SF}_{\infty}$. This is a remarkable result given that the \dm\ 
  distribution of every individual quasar is Gaussian. 
\item We made predictions for the incidence of quasar
  contamination in transient surveys using detailed simulations of
  quasar light curves from a mock LSST catalog. Due to the exponential
  nature of the \dm\ distributions for quasars, the probability of
  observing $\dm>1$~mag reaches 0.02 in the $u$ band (where
  variability is strongest) for time lags of 300~days, and $6\times
  10^{-4}$ for $\dm>2$~mag. Assuming Gaussian \dm\ distributions will
  result in erroneous likelihood estimates that are about 10 and 1000
  times smaller, respectively. 
\end{enumerate}

It is clear that a major limitation for the S82 quasars is the quality 
of light curves in both sampling density and time span. It is also
clear that our variability model needs to be better tested given the
evidence for a likely bias in the S82 time scale estimates. 
The best current sample for these improvements is that from the OGLE
microlensing survey, since the light curves are more densely-sampled
and longer than for S82. Here, the probem is the lack of spectroscopic 
identification of quasar candidates, although the follow-up
confirmation of quasars is rapidly improving (Koz{\l}owski et al.\ 2011b). 
The next-generation surveys will also greatly improve the constraints
on the long-term SF both individually and for ensembles of
quasars. The best short-term prospects are PTF (Law et al.\ 2009),
Pan-STARRS (Kaiser et al.~2002), and the Dark Energy Survey (DES;
Honscheid et al.~2008). In particular, the DES supernova program will
greatly expand many of the S82 quasar light curves with $griz$ 
sampling once per week for $\sim$3~months per year over 5~years.
The combination of SDSS, Pan-STARRS, DES, and LSST will yield
well-sampled light curves covering over 25 years for 10,000 quasars in
Stripe 82.    

The success of the model presented here suggests that a range of 
characteristic time scales exists among an ensemble of quasars, 
which can be related to physical time scales in the accretion disk. 
While we assumed a single $\tau$ per quasar, there is evidence 
that multiple time scales can exist for a given quasar (Collier 
\& Peterson 2001; Kelly et al.\ 2011). Therefore, the study presented
here can be extended to adopt the model in Kelly et al.\ (2011), which
fits more than one $\tau$ for a given object. This model, also called
a mixed Ornstein-Uhlenbeck (OU) process, reproduces a PSD of the form
exhibited by the X-ray light curves of galactic black holes and AGN,
which is flat below a low-frequency break, decays as $1/f$ above the
low-frequency break, and steepens to $1/f^2$ above a high-frequency
break.  In this case, with two characteristic time scales for each
quasar, the long-term ensemble SF can be revisited and possibly
explained in the context of a mixed OU process.  Note that recent
optical data from the Kepler mission (Mushotzky et al.\ 2011), which
have a sampling of 1 data point roughly every 30 minutes and 0.1\%
errors, suggest an additional break to a steeper slope ($\sim 1/f^3$),
but this dependence is seen on time scales shorter than can be
resolved in SDSS data.   

\vskip 0.4in \leftline{Acknowledgments}

We acknowledge support by NSF grant AST-0807500 to the University of
Washington, and NSF grant AST-0551161 to LSST for design and
development activity. CSK is supported by NSF grant AST-1009756. We
thank the referee for a thorough review and suggestions which led to an improved manuscript.  

    Funding for the SDSS and SDSS-II has been provided by the Alfred P. Sloan Foundation, the Participating Institutions, the National Science Foundation, the U.S. Department of Energy, the National Aeronautics and Space Administration, the Japanese Monbukagakusho, the Max Planck Society, and the Higher Education Funding Council for England. The SDSS Web Site is http://www.sdss.org/.

    The SDSS is managed by the Astrophysical Research Consortium for the Participating Institutions. The Participating Institutions are the American Museum of Natural History, Astrophysical Institute Potsdam, University of Basel, University of Cambridge, Case Western Reserve University, University of Chicago, Drexel University, Fermilab, the Institute for Advanced Study, the Japan Participation Group, Johns Hopkins University, the Joint Institute for Nuclear Astrophysics, the Kavli Institute for Particle Astrophysics and Cosmology, the Korean Scientist Group, the Chinese Academy of Sciences (LAMOST), Los Alamos National Laboratory, the Max-Planck-Institute for Astronomy (MPIA), the Max-Planck-Institute for Astrophysics (MPA), New Mexico State University, Ohio State University, University of Pittsburgh, University of Portsmouth, Princeton University, the United States Naval Observatory, and the University of Washington.

\appendix{\bf Appendix}

Here we present a database of \about 3.5 million photometric measurements
for 80,000 spectroscopically confirmed quasars to be used for time
variability studies. The database consists of three different data
sets: two with repeated SDSS imaging in five UV-to-IR photometric
bands, and one with SDSS versus POSS imaging for three bands. The
observed time lags span the range from 0.8 days to 10 years 
for the SDSS data sets, and up to 20 years for the SDSS vs.\ DPOSS data
set. The three data sets are described below: 
\begin{enumerate}
\item \emph{Northern Survey:} The SDSS imaging data are obtained by drift
  scanning. Because of the scan overlaps, and because of the scan
  convergence near the survey poles, about 40\% of the northern survey
  area (\about 4000 deg$^2$) is surveyed at least twice. This method
  provides 2-epoch 5-band coverage for \about 25,000 spectroscopically
  confirmed quasars. We adopt the SDSS BEST photometry listed in the
  DR7 Quasar Catalog V (Schneider et al.\ 2010) for the primary
  observations, and we searched for all unresolved secondary observations within
  1 arcsecond of the primary using CasJobs. Here, we include all
  observations up to three per quasar (only 0.2\% of the sample had
  more than three observations). We also include the redshifts,
  absolute magnitudes, FIRST, RASS, and 2MASS photometry as listed in
  the DR7 Quasar Catalog V, and the black hole masses as measured from
  emission line widths by Shen et al.\ (2011).    
  The catalog format is found in Table~\ref{tab:northerndmformat}. We
  also provide a list of all the objects (in the same order) with the
  same exact format as the DR7 Quasar Catalog V (Table~2 in Schneider
  et al.\ 2010).   
\item \emph{Southern Survey:} About 290 square degrees of the southern
  survey area has already been 
  observed \about 60 times to search for variable objects and, by
  stacking the frames, to go deeper. This is the SDSS Stripe 82, which
  is 22h 24m $<$ R.A.\ $<$ 04h 08m and $|{\rm Dec}| < 1.27$ deg. These
  multi-epoch data have time scales ranging from 3 hours to almost 10
  years. This method provides well-sampled 5-band light curves for an
  unprecedented number of quasars (9,258). 
  The catalog format is found in Table~\ref{tab:southerndbformat}, and
  the light curve file format is found in Table~\ref{tab:southernlcformat}. We
  also provide a list of all the objects (in the same order) with the
  same exact format as the DR5 Quasar Catalog IV (Table~2 in Schneider
  et al.\ 2007). 
\item \emph{SDSS--DPOSS:}  We also include a catalog of all SDSS DR7 quasars 
  with DPOSS observations. Following the procedure outlined in Sesar et al.\ (2006), 
  we have recalibrated DPOSS data (Djorgovski et al.\ 1998) in 8,000 deg$^2$
  of sky from the SDSS Data Release 5. The main advantage of this data
  set, which includes 81,189 quasars, is its long time
  baseline of 20 years. Here, we present the SDSS--DPOSS photometry in
  $GRI$ bands, where the latter is accurate to 0.10--0.15 mag. The
  catalog format is presented in Table~\ref{tab:sdsspossformat}.
\end{enumerate}

For more details, see \\
\scriptsize http://www.astro.washington.edu/users/ivezic/macleod/qso\_dr7/. 
\normalsize

\begin{deluxetable}{lrrl}
\footnotesize
\tablenum{2} \tablecolumns{4} \tablewidth{500pt}
\tablecaption{Northern Survey: Catalog Format \label{tab:northerndmformat}}
\tablehead {Column & Format & Units &  Description}
\startdata
1 & I6   & --- & Row of DR7 Quasar Catalog V (out of 105783)  \\
2 & F7.3 & mag & Absolute $i$ band magnitude\tablenotemark{h} \\
3 & F6.4 & --- & Redshift \\
4 & F6.3 &[$M_{\odot}$]& Black hole mass (``0'' means none could be found) \tablenotemark{g}\\
5 & F6.3 &[ergs/s]& Bolometric luminosity\tablenotemark{g} \\
6 & F6.3 & mag &  Galactic extinction in u filter (from Schlegel et al.\ 1998)\tablenotemark{c} \\
7 & F7.3 & mag &  FIRST Peak 20 cm flux density\tablenotemark{d} \\ 
8 & F8.3 & --- &  Signal-to-noise ratio for FIRST flux density  \\
9 & F8.3 &[ct/s]& RASS BSC/FSC full band count rate\tablenotemark{e} \\
10& F7.3 & --- &  Signal-to-noise ratio for RASS count rate           \\
11& F7.3 & mag &  2MASS $J$ band magnitude\tablenotemark{f}    \\
12& F6.3 & mag &  Error in $J$			      \\
13& F7.3 & mag &  2MASS $H$ band magnitude\tablenotemark{f}   	\\
14& F6.3 & mag &  Error in $H$ 			      \\
15& F7.3 & mag &  2MASS $K$ band magnitude\tablenotemark{f}	\\
16& F6.3 & mag &  Error in $K$					      \\
17& I3   & --- &  Total number of observations            \\
18& I1   & --- &  Flag for Stripe 82 object\tablenotemark{a}  \\
19& F9.3 & days&  Modified Julian Date of imaging observation       \\
20& F6.3 & mag &  BEST SDSS $u$ band PSF magnitude\tablenotemark{b} \\
21& F6.3 & mag &  Error in $u$    			      	      \\
22& F6.3 & mag &  BEST SDSS $g$ band PSF magnitude\tablenotemark{b} \\
23& F6.3 & mag &  Error in $g$				      \\
24& F6.3 & mag &  BEST SDSS $r$ band PSF magnitude\tablenotemark{b} \\
25& F6.3 & mag &  Error in $r$				      \\
26& F6.3 & mag &  BEST SDSS $i$ band PSF magnitude\tablenotemark{b} \\
27& F6.3 & mag &  Error in $i$				      \\
28& F6.3 & mag &  BEST SDSS $z$ band PSF magnitude\tablenotemark{b} \\
29& F6.3 & mag &  Error in $z$                                    \\
...& ... & ... &  Fields 19 through 29 repeated for each
additional observation \\
\enddata
\tablenotetext{a}{\scriptsize A value of 1 (0) indicates a (non-)Stripe 82 source.} 
\tablenotetext{b}{\scriptsize SDSS photometric measurements are asinh magnitudes (Lupton, Gunn, 
  \& Szalay 1999) and are normalized (to \about 3\% accuracy) 
   to the AB-magnitude system (Oke \& Gunn 1983).  
   Uncorrected for Galactic extinction. A value of 0.000 indicates that the 
   value could not be retrieved from the SDSS database.} 
\tablenotetext{c}{\scriptsize Where $A_g$, $A_r$, $A_i$, $A_z = 0.736$, 0.534, 0.405, $0.287 \times A_u$, respectively.}
\tablenotetext{d}{\scriptsize In AB magnitudes; $-2.5 \log (f_{\nu} / 3631 {\rm Jy})$.}
\tablenotetext{e}{\scriptsize X-ray data from ROSAT All Sky Survey Bright and Faint source catalogs. 
  A value of $-9.000$ indicates a non-detection.}
\tablenotetext{f}{\scriptsize All 2MASS data are from the 2MASS All-Sky Data Release Point Source 
Catalog (PSC) as of 2003 March 25. Note that 2MASS measurements are Vega-based, not AB, magnitudes.  
A value of 0.000 indicates a non-detection. }
\tablenotetext{g}{\scriptsize Taken from the Shen et al.\ (2011) catalog.} 
\tablenotetext{h}{\scriptsize For $\Omega_m=0.300$, $\Omega_{\Lambda}=0.700$, $h=0.70$, and $\alpha_Q = -0.50$.}
\end{deluxetable}

\begin{deluxetable}{lrrl}
\footnotesize
\tablenum{3} \tablecolumns{4} \tablewidth{500pt}
\tablecaption{Southern Survey: Catalog Format \label{tab:southerndbformat}}
\tablehead {Column & Format & Units &  Description}
\startdata
1  & I7   & ---        & The name of the light curve file \\
2  &F10.6 & deg        & Median Right Ascension in decimal degrees (J2000)\\
3  &F10.6 & deg        & Median Declination in decimal degrees (J2000)\\
4  & I5   & ---        & Row of DR5 Quasar Catalog IV (out of 77429)  \\
5  & F7.3 & mag        & Absolute $i$-band magnitude, K-corrected to $z=0$\tablenotemark{d} \\
6  & F7.3 & mag        & Absolute $i$-band magnitude, K-corrected to $z=2$\tablenotemark{a} \\
7  & F6.4 & ---        & Redshift \\
8  & F5.3 &[M$_{\odot}$]& Black hole mass (``0.000'' means none could be found)\tablenotemark{a}\\
9  & F6.3 & [ergs/s]   & Bolometric luminosity\tablenotemark{a} \\
10 & F6.3 & mag        & SDSS BEST $u$-band PSF magnitude\tablenotemark{b}  \\
11 & F6.3 & mag        & SDSS BEST $g$-band PSF magnitude\tablenotemark{b}  \\
12 & F6.3 & mag        & SDSS BEST $r$-band PSF magnitude\tablenotemark{b}  \\
13 & F6.3 & mag        & SDSS BEST $i$-band PSF magnitude\tablenotemark{b}  \\
14 & F6.3 & mag        & SDSS BEST $z$-band PSF magnitude\tablenotemark{b}  \\
15 & F6.3 & mag        & Galactic extinction in $u$ filter (from Schlegel et al.\ 1998)\tablenotemark{c} \\
\enddata
\tablenotetext{a}{\scriptsize Taken from the Shen et al.\ (2008) catalog. A value of ``-1'' indicates a
newly confirmed DR7 quasar, and thus is not present in Shen et al.\ (2008)
(see the latest version, Shen et al.\ 2011, for the values). }
\tablenotetext{b}{\scriptsize SDSS photometric measurements are asinh magnitudes (Lupton, Gunn, 
  \& Szalay 1999) and are normalized (to \about 3\% accuracy) 
   to the AB-magnitude system (Oke \& Gunn 1983).  
   Uncorrected for Galactic extinction. A value of 0.000 indicates that the 
   value could not be retrieved from the SDSS database.} 
\tablenotetext{c}{\scriptsize Where $A_g$, $A_r$, $A_i$, $A_z = 0.736$, 0.534, 0.405, $0.287 \times A_u$, respectively. 
  This value is set to zero if it is a newly confirmed DR7 quasar (see the DR7 Quasar Catalog V for the true values).}
\tablenotetext{d}{\scriptsize For $\Omega_m=0.300$, $\Omega_{\Lambda}=0.700$, $h=0.70$, and $\alpha_Q = -0.50$.}
\end{deluxetable}

\begin{deluxetable}{lrrl}
\footnotesize
\tablenum{4} \tablecolumns{4} \tablewidth{500pt}
\tablecaption{Southern Survey: Light Curve File Format \label{tab:southernlcformat}}
\tablehead {Column & Format & Units &  Description}
\startdata
1  & D    & days & Modified Julian Date for $u$-band observation \\
2  & F6.3 & mag  & Apparent $u$-band magnitude\tablenotemark{a}\\
3  & F5.3 & mag  & Error in $u$    	  \\
4  & D    & days & Modified Julian Date for $g$-band observation \\
5  & F6.3 & mag  & Apparent $g$-band magnitude\tablenotemark{a}\\
6  & F5.3 & mag  & Error in $g$\\
7  & D    & days & Modified Julian Date for $r$-band observation \\
8  & F6.3 & mag  & Apparent $r$-band magnitude\tablenotemark{a}\\
9  & F5.3 & mag  & Error in $r$\\
10 & D    & days & Modified Julian Date for $i$-band observation \\
11 & F6.3 & mag  & Apparent $i$-band magnitude\tablenotemark{a}\\
12 & F5.3 & mag  & Error in $i$\\
13 & D    & days & Modified Julian Date for $z$-band observation \\
14 & F6.3 & mag  & Apparent $z$-band magnitude\tablenotemark{a}\\
15 & F5.3 & mag  & Error in $z$\\
16 & D    & deg  & Median Right Ascension in decimal degrees (J2000)\tablenotemark{b}\\
17 & D    & deg  & Median Declination in decimal degrees (J2000)\\
\enddata
\tablenotetext{a} {Not corrected for Galactic absorption. Bad observations
are printed as ``-99.99''.}
\tablenotetext{b} {360 deg is subtracted from all RA values exceeding 300 deg.}
\end{deluxetable}

\begin{deluxetable}{lrrl}
\footnotesize
\tablenum{5} \tablecolumns{4} \tablewidth{500pt}
\tablecaption{SDSS--DPOSS: Catalog Format \label{tab:sdsspossformat}}
\tablehead {Column & Format & Units &  Description}
\startdata
1  & F10.6 & deg & Right Ascension in decimal degrees (J2000)      \\
2  & F10.6 & deg & Declination in decimal degrees (J2000)      \\
3  & I3 & --- & DPOSS Plate ID\\
4  & F9.4 & yrs & Epoch of the DPOSS observation in the $G$ band \\ 
5  & F9.4 & yrs & Epoch of the DPOSS observation in the $R$ band \\ 
6  & F9.4 & yrs & Epoch of the DPOSS observation in the $I$ band \\
7  & F6.2 & mag & DPOSS catalog $G$ magnitude\tablenotemark{d}\\
8  & F6.2 & mag &      Error in $G$\\
9  & F6.2 & mag & $G$-band quality flag\tablenotemark{e}\\
10 & F6.2 & mag & DPOSS catalog $R$ magnitude\tablenotemark{d}\\
11 & F6.2 & mag &      Error in $R$\\
12 & F6.2 & mag & $R$-band quality flag\tablenotemark{e}\\
13 & F6.2 & mag & DPOSS catalog $I$ magnitude\tablenotemark{d}\\
14 & F6.2 & mag &      Error in $I$\\
15 & I1 & mag & $I$-band quality flag\tablenotemark{e}\\
16 & I6   & --- & Row of DR7 Quasar Catalog V (out of 105783) \\
17 & F7.3 & mag & Absolute $i$ band magnitude\tablenotemark{h} \\
18 & F6.4 & --- & redshift  \\
19 & F6.3 &[M$_{\odot}$]& Black hole mass (``0.000'' means none could be found)\tablenotemark{g}\\
20 & F6.3 &[ergs/s]    & Bolometric luminosity\tablenotemark{g} \\
21 & F6.3 & mag        & Galactic extinction in $u$ filter (from Schlegel et al.\ 1998)\tablenotemark{c} \\
22 & I1 & --- & Number of observations\\
23 & I1 & --- & Flag for Stripe 82 object\tablenotemark{a} \\
24 & I5  & days & Modified Julian Date for SDSS imaging observation \\
25 & F6.3 & mag & BEST SDSS $g$ band PSF magnitude\tablenotemark{b}\\
26 & F6.3 & mag & Error in $g$\\
27 & F6.3 & mag & BEST SDSS $r$ band PSF magnitude\tablenotemark{b}\\
28 & F6.3 & mag & Error in $r$\\
29 & F6.3 & mag & BEST SDSS $i$ band PSF magnitude\tablenotemark{b}\\
30 & F6.3 & mag & Error in $i$\\
\enddata
\tablenotetext{a}{\scriptsize A value of 1 (0) indicates a (non-)Stripe 82 source.} 
\tablenotetext{b}{\scriptsize SDSS photometric measurements are asinh magnitudes (Lupton, Gunn, 
  \& Szalay 1999) and are normalized (to \about 3\% accuracy) 
   to the AB-magnitude system (Oke \& Gunn 1983).  
   Uncorrected for Galactic extinction. A value of 0.000 indicates that the 
   value could not be retrieved from the SDSS database.} 
\tablenotetext{c}{\scriptsize Where $A_g$, $A_r$, $A_i$, $A_z = 0.736$, 0.534, 0.405, $0.287 \times A_u$, respectively.}
\tablenotetext{d}{\scriptsize Not corrected for Galactic absorption. Bad observations
are printed as ``-99.99''.}
\tablenotetext{e}{\scriptsize Flag indicating whether the recalibrated magnitude is 
  of good quality (a value of 1 is ``good''). See Ses06 for a description of these cuts.  }

\tablenotetext{g}{\scriptsize Taken from the Shen et al.\ (2011) catalog.}
\tablenotetext{h}{\scriptsize For $\Omega_m=0.300$, $\Omega_{\Lambda}=0.700$, $h=0.70$, and $\alpha_Q = -0.50$.}

\end{deluxetable}

\end{document}